\documentclass{ieeeaccess}
\usepackage{cite}
\usepackage{amsmath,amssymb,amsfonts}
\usepackage{algorithmic}
\usepackage{graphicx}
\usepackage{textcomp}

\usepackage{xcolor}

\usepackage{textcomp}
\usepackage{caption}
\usepackage{url}
\graphicspath{ {./imgs/} }
\usepackage{rotating}
\usepackage{multirow}
\usepackage{comment}
\usepackage{listings}

\def\BibTeX{{\rm B\kern-.05em{\sc i\kern-.025em b}\kern-.08em
    T\kern-.1667em\lower.7ex\hbox{E}\kern-.125emX}}
\begin{document}
\history{Preprint submitted to IEEE Access. Date of current version June 22, 2020.}
\doi{10.1109/ACCESS.2017.DOI}

\title{Security and Privacy for mHealth and uHealth Systems: a Systematic Mapping Study}
\author{\uppercase{Leonardo Horn Iwaya}\authorrefmark{1,2},
\uppercase{Aakash Ahmad\authorrefmark{3}, and \uppercase{M. Ali Babar}}.\authorrefmark{1,2}}
\address[1]{Centre for Research on Engineering Software Technologies, The University of Adelaide, Adelaide, SA, Australia}
\address[2]{Cyber Security Cooperative Research Centre (CSCRC), Australia}
\address[3]{College of Computer Science and Engineering, University of Ha'il, Saudi Arabia}
\tfootnote{The work has been supported by the Cyber Security Cooperative Research Centre (CSCRC) whose activities are partially funded by the Australian Government's Cooperative Research Centres Programme.}

\markboth
{Iwaya \headeretal: Security and Privacy for mHealth and uHealth Systems: a Systematic Mapping Study}
{Iwaya \headeretal: Security and Privacy for mHealth and uHealth Systems: a Systematic Mapping Study}

\corresp{Corresponding author: Leonardo Horn Iwaya (e-mail: leonardo.iwaya@adelaide.edu.au).}

\begin{abstract}
An increased adoption of mobile health (mHealth) and ubiquitous health (uHealth) systems empower users with handheld devices and embedded sensors for a broad range of healthcare services.
However, m/uHealth systems face significant challenges related to data security and privacy that must be addressed to increase the pervasiveness of such systems.
This study aims to systematically identify, classify, compare, and evaluate state-of-the-art on security and privacy of m/uHealth systems. 
We conducted a systematic mapping study (SMS) based on 365 qualitatively selected studies to (i) classify the types, frequency, and demography of published research and (ii) synthesize and categorize research themes, (iii) recurring challenges, (iv) prominent solutions (i.e., research outcomes) and their (v) reported evaluations (i.e., practical validations).
Results suggest that the existing research on security and privacy of m/uHealth systems primarily focuses on \textit{select group of control families} (compliant with NIST800-53), \textit{protection of systems and information}, \textit{access control}, \textit{authentication}, \textit{individual participation}, and \textit{privacy authorisation}.
In contrast, areas of \textit{data governance}, \textit{security and privacy policies}, and \textit{program management} are under-represented, although these are critical to most of the organizations that employ m/uHealth systems. 
Most research proposes new solutions with limited validation, reflecting a lack of evaluation of security and privacy of m/uHealth in the real world.
Empirical research, development, and validation of m/uHealth security and privacy is still incipient, which may discourage practitioners from readily adopting solutions from the literature. 
This SMS facilitates knowledge transfer, enabling researchers and practitioners to engineer security and privacy for emerging and next generation of m/uHealth systems.
\end{abstract}

\begin{keywords}
Security, privacy, mobile computing, ubiquitous computing, medical information systems, health information management, reviews
\end{keywords}

\titlepgskip=-15pt

\maketitle

\section{Introduction}
\label{sec:introduction}
\PARstart{S}{mart} systems and infrastructures rely on mobile and pervasive technologies to offer end-users with portable and context-sensitive services that range from social networking, mobile commerce, to smart and connected health care \cite{cmatos2014smart, sassone2016smart}. 
Considering service-driven computing for smart systems, the future of smart healthcare is hyper-connected, highly pervasive, and personalized \cite{ivanovi2017personalized}. 
Mobile and pervasive technologies for mobile health (mHealth) and ubiquitous health (uHealth) systems provide a wide range of wellness and fitness applications as well as clinical and medical systems \cite{who2011mhealth, milovsevic2011applications}.
m/uHealth fitness applications (apps for short) and medical systems impact activities and practices of individuals, patients, medical professionals, and health service providers \cite{beratarrechea2014impact}. 
Central to this technological revolution for m/uHealth systems -- providing smart and connected health care -- is context-sensitive information and health critical data. 
A typical example of this is an individual's diet and exercise routine (e.g., context-sensitive information) and its impact on person's health such as blood pressure, body weight and any disease (e.g., health critical data). 
It has been widely recognized that healthcare data is one of the most valuable assets to the health services providers and medical/health technologies (MedTech) companies \cite{tanner2017our}. 
m/uHealth apps and systems have proven to be critically important for collecting, processing, and analyzing health data to generate actionable insights for all the stakeholders. 
However, there are increasing concerns and challenges for the security and privacy of the data gathered, processed and stored by m/uHealth apps and systems \cite{blobel2016patient, negash2018constrained}.
A recently study by Gartner highlights the importance of human and technological aspects of security and risk management for privacy and security practices of healthcare service providers \cite{pessin2017top}.
Security and privacy breaches in healthcare information systems have serious negative impacts to its data subjects. 
Such impacts can range from embarrassment and reputation damage to various forms of discrimination that adversely affects the rights and freedoms as well as physical and mental health of individuals \cite{fung2006impact, iwaya2016secure, iyengar2018healthcare}.

Given that security and privacy concerns have emerged as the most challenging aspects for healthcare information systems, there is an urgent need to fully understand and address the security and privacy issues of m/uHealth apps from software system' lifecycle perspective \cite{iwaya2019engineering}.
The lifecycle includes but is not limited to requirement analysis, design, implementation, testing, and deployment of m/uHealth systems.
With an increasing trend to provide health services through mobile/ubiquitous technologies, there is a growing body of research on identifying challenges, proposing solutions, and highlighting open issues related to security and privacy aspects of m/uHealth systems \cite{algarni2019survey}. 
For example, researchers have proposed architectures \cite{tamizharasi2017iot}, implemented algorithms  \cite{xu2017mcc} and mechanisms to establish infrastructures \cite{xu2017mcc} for addressing the security and privacy of healthcare systems. 
An ever increasing number of healthcare systems are being developed by adopting security engineering practices and recommendations provided by the relevant agencies such as \cite{enisa2018ict}.

Given the growing body of published research on security and privacy of m/uHealth systems, there are increasing needs and opportunities to carry out secondary studies to consolidate the knowledge and evidence being produced for ease of access for practitioners and researchers. 
That is why a number of researchers have surveyed the literature on security and privacy for mHealth (e.g., \cite{abu2017mhealth, koffi2018mobile, algarni2019survey}). 
However, the existing reviews tend to limit the scope to mHealth systems (using smartphones, tablets and wearable sensors), but they do not explicitly include more pervasive and context-sensitive uHealth systems. 
Moreover, the existing reviews have focused on very specific security controls used in m/uHealth systems, such as biometrics, authentication, and key exchange schemes \cite{mohsin2018biometric, shuwandy2019authentication}. 
Hence, the existing research and specifically survey based studies (detailed in a dedicated section) lack a broader view on the topic, i.e., explicitly comprising both mHealth and uHealth as well as dealing with security and privacy controls related issues for the class of m/uHealth systems. 
To support the engineering of m/uHealth systems with security and privacy aspects embedded right from
the beginning \cite{tamizharasi2017iot, xu2017mcc}, there is a need to systematically select, review, and synthesize the published research on security and privacy of m/uHealth systems. 
Review and synthesis of published research helps to classify, compare, and evaluate the strengths and limitations of the state-of-the-art in the area under investigation.

To address the above goals, we have carried out a Systematic Mapping Study (SMS) \cite{petersen2008systematic} of the peer-reviewed literature on security and privacy for m/uHealth systems. 
To complement the SMS, we also performed an in-depth thematic analysis of the studies that have been evaluated in practice, discussing the reported solutions, their evaluation strategies and the impacts on the industry scale systems. 
We are not aware of any other effort that has carried out a SMS for identifying, classifying, comparing, and communicating the existing research and its implications to the relevant stakeholders (i.e., researchers, practitioners, policy-makers, healthcare providers, and broader society). 
Hence, this SMS provides an overview of the topic in terms of: 
1) research and contribution types; 
2) research trends and taxonomy; 
3) challenges and solutions for security and privacy controls; 
5) m/uHealth application categories; and, 
6) role of various devices and technologies in m/uHealth systems. 
Bibliographical information and trends of research also pinpoint the predominant areas of research, under-researched areas and gaps, as well as the future research directions.

The core results of this SMS highlight that the existing research on m/uHealth system frequently emphasises the use of security and privacy controls of a small group of families, considering the NIST 800-53 control families. 
The predominant research trends reflect: 1) system and communication protection, 2) identification and authentication, 3) system and information integrity, 4) access control, 5) individual participation and 6) privacy authorisation. 
Researchers are mostly focused on investigating most traditional families of security and privacy controls, however, other areas such as data governance, security and privacy policies and program management remain under-investigated even though they are crucial for dealing with security and privacy at an organizational level. 
The primary contributions of this SMS are:  
\begin{itemize}
    \item Classification and comparison of the existing and emerging solutions for security and privacy for m/uHealth in the form of systematic maps, classification taxonomy, and illustrative trends. 
    \item Evaluation focused analysis of the solutions - implemented in practice - to identify commons themes and appraising the quality of these evaluation studies.
\end{itemize}

Empirical evidence along with research and development of security and privacy solutions is lacking and research studies need to be carefully evaluated before academic solutions can be adopted or extended in an industrial context. 
The results of this SMS can be beneficial for:
\begin{itemize}
    \item Researchers who are interested in quickly identifying the existing research that can help to formulate new hypothesis to be tested and propose innovative solutions for the emerging challenges of security and privacy of m/uHealth systems.
    \item Practitioners who want to understand the solutions reported in the academic literature to provide architectural models, implementation strategies and evaluation frameworks that can be evaluated for industrial adoption. 
\end{itemize}

This papers is structured as follows.
Section \ref{sec:background} discusses context and background details for security and privacy of m/uHealth systems. 
Section \ref{sec:related-work} reviews related surveys to justify the scope and contribution of the proposed SMS. 
Section \ref{sec:methodology} presents research methodology. 
Section \ref{sec:results-part1}, Section \ref{sec:results-part2}, and Section \ref{sec:results-part3} discuss the core results of the SMS. 
Section \ref{sec:discussion} reviews the results to present the critical findings of the SMS. 
Section \ref{sec:threats-to-validity} presents validity threats to the SMS. 
Section \ref{sec:conclusion} concludes the paper. 

\section{Background}
\label{sec:background}
In this section, we contextualize the security and privacy issues of m/uHealth systems, as illustrated in Figure \ref{fig:background}. 
First, we discuss mHealth and uHealth in the context of electronic healthcare (eHealth) systems (Section \ref{sec:back-health-it}). 
Second, we conceptualize security and privacy, their interrelations, standards and legal frameworks (Section \ref{sec:back-sec-pri}). 
The concepts and terminologies introduced in this section are used throughout the paper.

\subsection{eHealth, mHealth and uHealth}
\label{sec:back-health-it}
\begin{figure*}[h]
  \centering
    \includegraphics[width=0.6\textwidth]{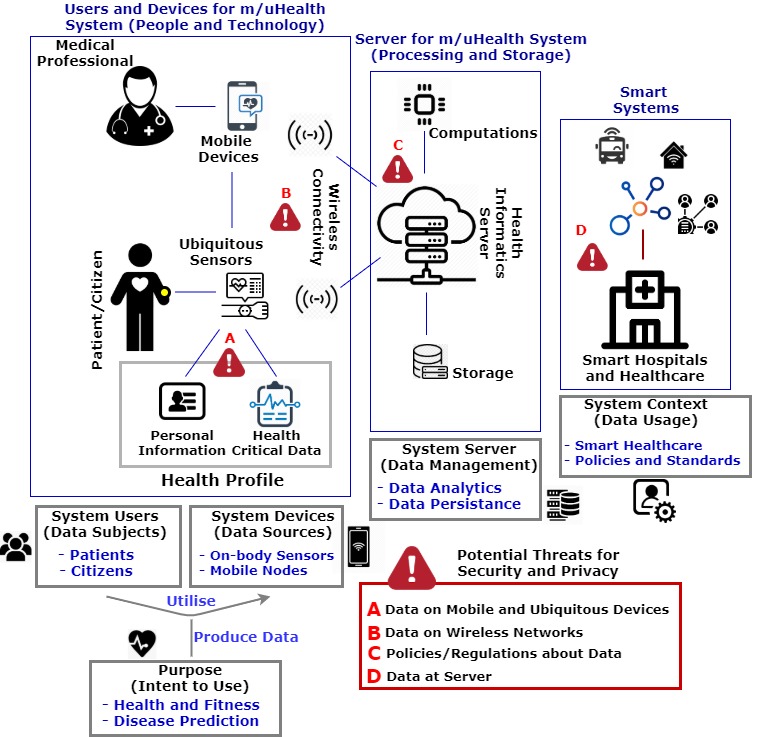}
  \caption{Overview of m/uHealth Systems in the context of Security and Privacy Issues.}
  \label{fig:background}
\end{figure*}

Since the late 90's, the term of electronic health (eHealth) is used to refer to healthcare processes and practices supported by Information and Communication Technologies (ICT) \cite{adibi2015mhealth}. 
There are various forms of eHealth systems, such as telemedicine, Electronic Health Records (EHR) and Healthcare Information Systems. 
mHealth is the practice of eHealth assisted by smartphones and other mobile devices, used to collect, analyze, process, transmit and store health services related information from sensors and other biomedical systems \cite{adibi2015mhealth}, as in Figure \ref{fig:background}. 
The paradigm of uHealth -- driven by ubiquitous computing\footnote{Ubiquitous computing was introduced by Weiser \cite{weiser1999ubicomp}, as \textit{``the integration of computers seamlessly into the world''}.} -- is as an advancement of mHealth systems that exploit ubiquitous devices and sensors to enable on-the-go health monitoring and care \cite{milovsevic2011applications}.
uHealth is fast emerging as a pervasive technology that uses a large range of sensors and actuators deployed in an environment (e.g., homes, hospitals or workplaces) or used by individuals (e.g., worn/implanted on-body sensors) to support  the delivery of healthcare, monitoring, and improving individuals' physical and mental health \cite{brown2007ethical}. 
As in Figure \ref{fig:background}, m/uHealth systems empower their users with anytime/anywhere sensors, applications and networks that collect context-sensitive health critical data in the form of blood pressure, heart rate and body temperature to diagnose health related problems \cite{brown2007ethical}. 
As illustrated in Figure \ref{fig:background}, user/citizes/patients can exploit their on-body sensors that can monitor their health related information (a.k.a. medical profiles) that can be shared with medical professionals at a distributed locations. 
Medical professionals can use their mobile devices for medical consultation and data stored on health care servers could be shared with other professionals or health care units. 
Despite the offered benefits by m/uHealth systems, one of the most critical challenges for this class of systems relates to security and privacy of an individual's personal data and health critical information \cite{iwaya2019engineering}.

\subsection{Security and privacy}
\label{sec:back-sec-pri}
In a healthcare environment, security and privacy of information systems is critical for achieving trust and high-quality services \cite{iwaya2019engineering}. 
Although the terms security and privacy tend to be used interchangeably across research, the concepts have in fact fundamental differences that should be taken into account when dealing with m/uHealth systems. 
In general, security of computing systems targets the protection and safeguarding of hardware, software and information, and typically boils down to three core concepts \cite{bishop2005introduction}: 
(1) confidentiality, the concealment of information or resources; 
(2) integrity, the trustworthiness of data resources; prevention of improper or unauthorized change; and, %
(3) availability, the ability to use the  information or resource desired. 
Privacy, in turn is not simply a technical concept. 
It is a fundamental human right, both in terms of physical privacy and information privacy \cite{diggelmann2014privacy}.
Here we are particularly interested in the latter, i.e., privacy in the context of health critical data being produced and consumed by m/uHealth systems. 
Although there is no absolute agreement on the definition of privacy, in this paper we consider the proposition of Westin \cite{westin2015privacy}: 
\textit{``Privacy is the claim of individuals, groups or institutions to determine for themselves when, how and to what extent information about them is communicated to others.''}
Informational privacy overlaps with the concept of confidentiality, such as the authorized access or disclosure of information and notions of secrecy, access-control, sharing and protection of information \cite{iwaya2019engineering, danezis2010critical}. 
However, in order to create privacy-aware systems other aspects such as openness and transparency, purpose specification and limitation and informed consent also need to be put in place. 
Although computer security is an essential pillar for achieving privacy, it should be clear that privacy cannot be satisfied solely on the basis of managing security \cite{nist8062}.

\subsection{Security and privacy in the context of m/uHealth Systems}
As illustrated in Figure \ref{fig:background}, per the model and guidelines for secure mobile and ubiquitous systems from \cite{sajjad2018mobcomp}, we highlight that the security and privacy issues for health critical data can be categorized into four main categories. 
These include security and privacy of (1) data produced and consumed by mobile and ubiquitous devices, (2) data transmitted over wireless networks, (3) information  residing on healthcare servers, and (4) policies and regulations for m/uHealth systems (detailed later).
For example, as in Figure \ref{fig:background} shows a user can exploit his/her body sensors to collect health critical data such as heartbeat, body temperature, and blood pressure referred to as Health Profile. 
The Health Profile is composed of personal information (e.g., age, gender or location) and health critical data (e.g., blood pressure, body temperature or pulse rate). 
Consider Figure \ref{fig:background}, a typical scenario can be a compromised device or sensor for unauthorized access to user's health profile with an intent for delivering personalized advertisement, or leakage of personal information for social profiling.. 
Such and alike scenarios that breach security and privacy protocols of m/uHealth systems can limit users' trust and adoption of such systems \cite{abu2017mhealth, koffi2018mobile}.

\subsection{Policies and regulations for security and privacy}
In the past years, some regulations have been enacted such as the European General Data Protection Regulation (EU GDPR) \cite{GDPR2016} and the California Consumer Privacy Act (CCPA) \cite{CCPA2018}, giving individuals more power over their data and putting more limitations on the ways data is collected, analyzed and used by organisations. 
At the same time, standardisation bodies have been providing guidance to systems architects in terms of engineering secure and privacy-preserving systems. 
Examples are the NIST's revised list of security and privacy controls (NIST 800-53~\footnote{https://csrc.nist.gov/publications/detail/sp/800-53/rev-5/draft}) and Privacy Engineering Program (PEP~\footnote{https://www.nist.gov/itl/applied-cybersecurity/privacy-engineering}), and the ISO's standard on privacy engineering for system life cycle processes (ISO/IEC TR 27550:2019~\footnote{https://www.iso.org/standard/72024.html}). 
Achieving compliance and implementing standards is particularly interesting in the context of mHealth and uHealth systems, because they are relatively new technologies and the security and privacy challenges have not been fully understood yet \cite{stewart1999privacy, fridewald2010privacy}. 
This discussion is also elaborated in this paper, putting in perspective the current advances of m/uHealth and recent changes in the context of policies and regulations for security and privacy of m/uHealth systems.

\section{The existing surveys and systematic reviews on security and privacy of m/uHeatlh}
\label{sec:related-work}
In this section, we review the most relevant survey-based (secondary research) studies that focus on privacy and usability, security, and other relevant aspects for health critical software systems with a particular focus on m/uHealth. 
We justify the scope and contributions of the proposed SMS based on a systematic comparison of the existing survey based studies in Table \ref{tab:comp-surveys}.

\begin{table*}
\caption{Comparison of existing surveys and SLRs for m/uHealth systems.}
\label{tab:comp-surveys}
\begin{tabular}{|p{0.03\textwidth}|p{0.06\textwidth}|p{0.03\textwidth}|p{0.06\textwidth}|p{0.07\textwidth}|p{0.03\textwidth}|p{0.3\textwidth}|p{0.25\textwidth}|}
\hline
\textbf{Ref} & \textbf{Pub. Type} & \textbf{Pub. Year} & \textbf{Reviewed Studies} & \textbf{Coverage Years} & \textbf{SLR} & \textbf{Focus} & \textbf{Limitation}\\ \hline \hline
\multicolumn{8}{|c|}{\textbf{Assessment of mHealth apps}} \\ \hline
\cite{martinez-perez2014review} & Journal & 2015 & 169 & 2007-2014 & Yes & assessment of  mHealth apps & Not covering uHealth \\
\cite{scott2015review} & Journal & 2015 & N.A. & N.A. & No & security and privacy of  mHealth apps & \\
\cite{abu2017mhealth} & Journal & 2017 & 7 & 2011-2016 & Yes & mHealth apps and user's perspectives & \\
\cite{koffi2018mobile} & Conf. & 2018 & 37 & 2011-2016 & Yes & privacy concerns of patients and providers & \\
\cite{househ2018balancing} & Book Ch. & 2018 & 68 & N.A.-2017 & Yes* & trade-off of privacy and patients needs for data & \\
\cite{katusiime2019review} & Journal & 2017 & 22 & 2001-2016 & Yes & privacy and usability issues in mHealth systems & \\
\cite{al12networking} & Journal & 2019 & 201 & 2007-2014 & Yes & assessment of mHealth apps & \\ \hline \hline
\multicolumn{8}{|c|}{\textbf{Authentication for m/uHealth systems}} \\ \hline
\cite{mohsin2018biometric} & Journal & 2018 & 150 (19) & 2012-2017 & Yes & telemedicine sensors; finger vein biometrics & Not fully covering security nor privacy\\
\cite{shuwandy2019authentication} & Journal & 2018 & 150 (19) & 2012-2017 & Yes & telemedicine sensors; sensor authentication & \\ \hline \hline
\multicolumn{8}{|c|}{\textbf{Security of implantable m/uHealth systems}} \\ \hline
\cite{camara2015implant} & Journal & 2015 & N.A. & N.A. & No & security of implantable medical devices & Limited to implantable devices \\ \hline \hline
\multicolumn{8}{|c|}{\textbf{Security of WSNs and WBANs}} \\ \hline
\cite{kumar2012security} & Journal & 2012 & N.A. & N.A. & No & issues in WMSNs & Limited to sensors-based m/uHealth \\
\cite{latif2014cloud-wban} & Journal & 2014 & 24 (7) & 2009-2014 & Yes & DDoS attack in cloud-assisted WBANs & \\
\cite{chukwunonyerem2014wban} & Conf. & 2014 & N.A. & N.A. & No & WBANs & \\
\cite{kang2015wban} & Conf. & 2015 & N.A. & N.A. & No & protocols on mHealth WBANs & \\
\cite{meharouech2019bsn} & Journal & 2019 & N.A. & N.A. & No & body-to-body sensor networks for uHealth & \\ \hline \hline
\multicolumn{8}{|c|}{\textbf{Security of IoT-based systems}} \\ \hline
\cite{somasundaram2017review} & Book Ch. & 2017 & N.A. & N.A. & No & IoT medical devices communication & Limited to IoT-based m/uHealth systems \\
\cite{ahmed2018insider} & Journal & 2018 & 21 (10) & 2009-2017 & Yes & malicious insiders in IoT/cloud-based eHealth & \\
\cite{hatzivasilis2019iomt} & Conf. & 2019 & N.A. & N.A. & No & security and privacy for the IoMT & \\ \hline \hline
\multicolumn{8}{|c|}{\textbf{Security of cloud-based e/mHealth systems}} \\ \hline
\cite{rodrigues2013cloud} & Journal & 2013 & N.A. & N.A. & No & security and privacy of cloud-based EHRs & Limited to cloud-based eHealth \\
\cite{wang2019mcc} & Journal & 2019 & N.A. & N.A. & No & security and privacy of MCC for healthcare & \\ \hline \hline
\multicolumn{8}{|c|}{\textbf{Broader ethical discussions}} \\ \hline
\cite{sharp2017ethics} & Conf. & 2017 & 84 & 2012-N.A. & Yes & ethics of mobile medical apps and mHealth & Emphasis on ethics, not security and privacy \\
\cite{maher2019ethics} & Journal & 2019 & 48 & N.A.-2018 & Yes & ethics on passive data collection in healthcare & \\ \hline \hline
\multicolumn{2}{|c|}{\textbf{Proposed SMS}} & -- & 365 & 2015-2019 & Yes & mHealth and uHealth; security and privacy & Not covering usability and ethics \\ \hline
\end{tabular}
\end{table*}

\subsection{Privacy and usability issues in mHealth systems}
Security and privacy of mHealth ``apps'' is among the most focused research domain in the context of secure and private mobile and ubiquitous systems. 
Researchers usually investigate the existing legal frameworks and academic literature in the area of mHealth in order to provide guidelines and recommendations to users (e.g., patients, clinicians or administrative people) and developers of mHeath apps \cite{martinez-perez2014review, abu2017mhealth, koffi2018mobile, househ2018balancing}. 
Some researchers have also defined a criteria for assessing and comparing mHealth apps, providing a security and privacy ``score'' in order to better communicate the potential risks of such systems to its end-users \cite{scott2015review}. 
Some systematic reviews have focused on privacy and usability issues in mHealth systems \cite{katusiime2019review} or networking issues for security and privacy of mHealth apps \cite{al12networking}. 
However, these studies are restricted to the investigation of mHealth apps developed for tablets and smartphones and Body Sensor Networks (BSNs). 
That is, the existing literature explicitly focuses on mHealth systems but falls short of addressing security and privacy issues for uHealth systems, which are being increasingly adopted in smart healthcare that is driven by IoTs and sensors-driven technologies \cite{tamizharasi2017iot}.

\subsection{Security in telemedicine and mHealth systems}
Considering telemedicine as an essential component of eHealth systems, some systematic reviews have focused security controls for such systems \cite{who2011mhealth}. 
A typical example of this is multi-layer systematic reviews for user authentication in telemedicine and mHealth systems, using body sensor information and finger vein biometric verification \cite{mohsin2018biometric} and sensor-based smart phones \cite{shuwandy2019authentication}. 
In such reviews, the prime focus of the investigation is authentication and key exchange protocols for secure telemedicine. 
These surveys review a myriad of schemes have been proposed for mHealth systems, typically addressing mutual authentication of users and application providers as well as for device authentication (e.g., mobile devices and sensors).

\subsection{Security and privacy of health critical systems}
In recent years, a number of systematic reviews and surveys have targeted security and privacy for health critical systems. 
For instance, the work of \cite{camara2015implant} focuses on security and privacy issues of implantable medical devices.
Also, a great deal of attention has been given to Wireless Body Area Networks (WBANs) in the context of mHealth \cite{chukwunonyerem2014wban, kang2015wban}. 
Such surveys have primarily focused on WBANs in terms of wireless medical sensor networks \cite{kumar2012security}, Distributed Denial of Service (DDoS) attacks \cite{latif2014cloud-wban} and body-to-body sensor networks \cite{meharouech2019bsn}. 
There are some reviews that have investigated Internet of Things (IoT) for medical applications \cite{somasundaram2017review, ahmed2018insider, hatzivasilis2019iomt}, as well as cloud-based EHRs \cite{rodrigues2013cloud} and Mobile Cloud Computing \cite{wang2019mcc}. 
These reviews represent a concentrated body of knowledge about the challenges and solutions on security for resource-constrained devices and cloud services, so as to ensure confidentiality of data in-transit (over network) and at-rest (residing in devices or servers). 
It is also worth mentioning that some systematic reviews go beyond the scope of security and privacy, addressing ethics altogether in the context of mHealth apps \cite{sharp2017ethics} and passive data collection in healthcare \cite{maher2019ethics}.

\textit{Conclusive summary:} 
We have discussed the progression and limitations of the existing research in terms of survey-based secondary studies that enable or enhance the security and privacy of m/uHealth systems. 
In order to justify the scope and contributions of the proposed SMS, an objective comparison is presented in Table \ref{tab:comp-surveys}. 
Table \ref{tab:comp-surveys} acts as a structured catalogue to classify and compare the existing research and the needs for futuristic research and development of security and privacy enabled m/uHealth. 
Unlike the existing surveys, the proposed SMS explicitly considers privacy controls, apart from the traditional security controls. 
This greater demand for privacy-by-design is in part a consequence of the recent privacy laws enacted in the European Union and United States (i.e., GDPR and CCPA). 
This SMS aims to address both security and privacy and explicitly includes uHealth as part of mHealth systems.

\section{Research method for SMS}
\label{sec:methodology}
\begin{figure*}
  \centering
    \includegraphics[width=0.7\textwidth]{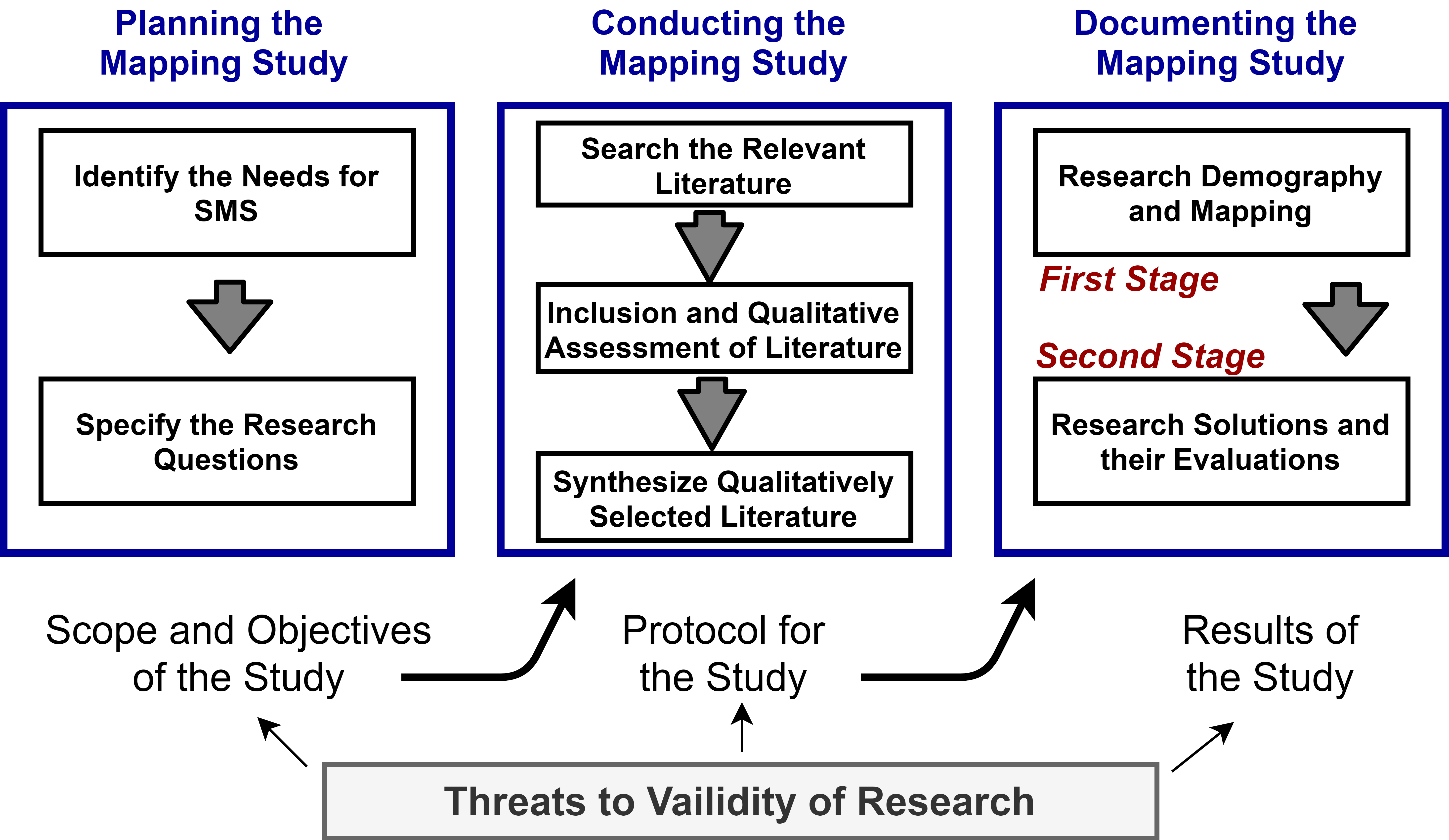}
  \caption{Methodology overview of the mapping process and evaluation focused analyses.}
  \label{fig:methodology}
\end{figure*}

In order to plan, conduct, and document this SMS, we followed the evidence based software engineering approach and specifically adhered to the guidelines to conduct systematic reviews and mapping studies from \cite{petersen2008systematic}.
We developed the research protocol to be followed at each step of this SMS. The details of the research protocol for this study are provided in \cite{iwaya2020protocol}). Hence, the full details of the research protocol for this study are not provided in this paper.
An illustrative view of the adopted research method is presented in Figure \ref{fig:methodology} that highlights three phases of research and each phase comprises of two tasks. 
Each phase has an outcome. 
For example, the  initial phase named Planning the Mapping Study comprises of two tasks that relate to (i) identification of the needs and (ii) specification of the research questions for the mapping study.
The outcome of this phase is scope and objectives of the SMS in terms of research questions that need to be investigated. 
SMS planning is the precondition for the later phases of the methodology. 
By adopting well-known methodology from \cite{petersen2008systematic} as in Figure \ref{fig:methodology}, we aim to strengthen the findings, support objective interpretation of results, minimize any bias, and enable reproducible results. 
In the remainder of this section, we discuss the three phases of the research methodology.

\subsection{Phase I -- planning a mapping study}
\label{sec:phase-i}
\subsubsection{Identify the needs for SMS}
\label{sec:identify-need}
Despite a lot of attention and published research, there is no effort to systematically identify and investigate a collective impact of the existing research on secure and private m/uHealth systems. 
A systematic investigation of the state-of-the-art on secure and private m/uHealth systems can highlight the research progression, maturation, emerging trends and futuristic challenges that are currently lacking in the literature. 
In order to ensure that no prior survey, mapping study, or systematic review (i.e., secondary studies) have been conducted, we searched the most prominent digital libraries including IEEE, ACM, Springer, Science Direct and Scopus along with indexing engine Google Scholar (search date 02/10/2019). 
The search string that we executed on these digital libraries and indexing engines to locate any secondary studies on security and privacy of m/uHealth systems is shown in Listing \ref{lst:search-string-slrs}. 
Based on Listing \ref{lst:search-string-slrs}, none of the retrieved literature (in. Section \ref{sec:background}, Table \ref{tab:comp-surveys}) was related to the outlined research questions in Section \ref{sec:research-questions} that motivated the need for this SMS.

\begin{center}
  \lstset{%
    caption=Search String to Identify Secondary Studies on Secure and Private m/uHealth.,
    basicstyle=\ttfamily\footnotesize\bfseries,
    frame=tb,
    backgroundcolor=\color{lightgray},
    label={lst:search-string-slrs},
    captionpos=b
  }
  \begin{lstlisting}
    (``Systematic Literature Review'' OR 
    ``Systematic Mapping'' OR ``SMS'' OR ``SLR'' 
    OR ``Study'' OR ``Survey'') AND 
    ((``ubiquitous health*'' OR ``uhealth'' OR 
    ``u-health'' ) OR (``mobile health*'' OR 
    ``mhealth'' OR ``m-health'') ) AND 
    (``secur*'' OR ``*security'' OR ``privacy*'' 
    OR ``crypto*'' )
  \end{lstlisting}
\end{center}

\subsubsection{Specifying the research questions}
\label{sec:research-questions}
To conduct the mapping study and present its results, we specify a number of Research Questions (RQs) for this SMS. 
The scope of this SMS is limited to finding and discussing the answers to the RQs that have motivated this study. 
The RQs and their respective objective(s) are described below:
\begin{enumerate}
    \item[\textbf{A.}] \textbf{Demography Analysis -- Types, Frequency, and Venues of Research Publications}
    \begin{enumerate}
        \item[RQ-1] \textit{What is the type of and frequency of the published research in the area of security and privacy of m/uHealth systems?} Objective(s): To understand and highlight the (i) type of published research in terms of conference proceedings, journal articles, symposium papers etc., and (ii) frequency of published research that can reflects the progress and growth of the research in terms of the number of publication over the years.    
        \item[RQ-2] \textit{What are the prominent venues of the research publications in the area under investigation?} Objective(s): To list and analyse the publication venues such as specific conference proceedings, journal articles and special issues along with book chapters that highlight the prominent venues of the research emergence that are detailed below.
    \end{enumerate}
    \item[\textbf{B.}] \textbf{Research Mapping -- Existing Solutions, their Evaluations, and Validation Research}
    \begin{enumerate}
        \item[RQ-3] \textit{What are the proposed solutions of security and privacy of m/uHealth systems?} Objective(s): Identification of the proposed solution represent various research themes -- reflecting the body of knowledge -- that helps us to investigate the strengths and limitations of the existing research.
        \item[RQ-4] \textit{What is the state of existing evaluation studies on security and privacy for m/uHealth systems?} Objective(s): Provide a clear picture of the existing research that has been properly evaluated. Evaluation of the existing research reflects the strength of the solutions in terms of their practical applicability and validation.
    \end{enumerate}
\end{enumerate}

\subsection{Phase II -- conducting a mapping study}
\label{sec:phase-ii}
As per Figure \ref{fig:methodology}, this phase involves searching and qualitative assessment of the relevant literature that is included for review to conduct the SMS, detailed as below.

\subsubsection{Search the relevant literature}
In order to search the relevant literature, we selected the Scopus digital library that indexes more than five thousand publishers, including highly relevant sources such as Elsevier, Springer, MEDLINE, EMBASE, IEEE Xplore and ACM. 
In order to search the relevant studies in Scopus, we considered the outlined RQs (from Section \ref{sec:identify-need}) to compose the search string based on the key terms that are presented in Table \ref{tab:key-terms}.

\begin{table}[h]
\caption{Key terms used divided by groups.}
\label{tab:key-terms}
\centering
\begin{tabular}{|c|c|c|}
\hline
\textbf{G1} & \textbf{G2} & \textbf{G3}\\\hline\hline
mobile health* & ubiquitous health* & *security\\
mHealth & uHealth & secur*\\
m-Health & u-Health & privacy*\\
 &  & crypto*\\\hline
\end{tabular}
\end{table}

We decided to limit the time period for our searches, a 5-year period (i.e., from year 2015 - 2019). 
A pilot search based on the search string in Listing \ref{lst:search-string} suggested there was little to no relevant publications on the topic before under investigation before the Year 2015. 
Therefore, in order to avoid an exhaustive search and minimize the risk of identifying a large number of irrelevant studies, we set the search criteria to only cover the literature from Year 2015 to 2019 that helped us to retrieve a total of 1249 potentially relevant publications. 
We also limited the search to peer reviewed scientific publications and book chapters that excludes any white papers, technical report or unpublished work.

\begin{center}
  \lstset{%
    caption=Composition of Search String for Literature Search.,
    basicstyle=\ttfamily\footnotesize\bfseries,
    frame=tb,
    backgroundcolor=\color{lightgray},
    label={lst:search-string},
    captionpos=b
  }
  \begin{lstlisting}
    TITLE-ABS-KEY(
        (("ubiquitous health*" OR "uhealth" OR 
        "u-health") OR ("mobile health*" OR 
        "mhealth" OR "m-health")) AND (secur* 
        OR *security OR privacy* OR crypto* )
    )
  \end{lstlisting}
\end{center}

\subsubsection{Study inclusion -- screening and qualitative assessment}
\label{sec:study-inclusion}
We followed a two step process of screening and qualitative assessment for selecting the most relevant publications for review out of the initial set of 1249 potentially relevant papers. We developed 
The details of the selection process can be found in the study's research protocol available at \cite{iwaya2020protocol}.

\subsubsection{Synthesize qualitatively the selected literature}
\label{sec:synthesize-literature}
The last task of the Phase II is the classification of the studies using systematic maps.
We also conducted an evaluation and focused analysis of the papers that have been implemented in practice.
The construction of the SMS's classification facets as well as the detailed explanation of the qualitative assessment of the evaluation studies can be found in the research protocol \cite{iwaya2020protocol}.

\subsection{Phase III -- documenting a mapping study}
\label{sec:phase-iii}
As per Figure \ref{fig:methodology}, the last phase of the SMS, i.e., \textbf{Documenting the Mapping Study} is detailed in the remainder of this paper. 
The results documentation is based on investigating the RQs (in Section \ref{sec:phase-i}) and presenting their findings. 
The results documentation is classified as (a) \textit{Research Demography and Mapping} (as First Stage in Section \ref{sec:results-part1}) and (b) \textit{Research Solutions and their Evaluations} (as Second Stage in Sections \ref{sec:results-part2} and \ref{sec:results-part3}). 
As part of the documentation, the critical finding of the SMS are reviewed and validity threats to the SMS results are also presented (in Section \ref{sec:discussion} and Section \ref{sec:threats-to-validity} respectively).
The artefacts used and created in this study are publicly available in a replication package, which can be found at \cite{github}.
The replication package includes the database search queries, the answer sets of these queries and the maps of the categorised papers and patterns.

\section{Demography analysis based on types, frequency and venues of the publications}
\label{sec:results-part1}
In this section, we answer RQ-1 and RQ-2 that focus on demography analysis of published research on security and privacy of m/uHealth systems. 
In the demography analysis we aim to investigate the types of published research and frequency of publications during a specific year or a range of years (in Section \ref{sec:freq-types-research}). 
We also identify prominent venues of publications (Section \ref{sec:prominent-venues}). 
As per the research method (in Figure \ref{fig:methodology}), demography analysis relates to first stage analysis and documentation of the topic under investigation. 
Specifically, the Stage 1 mapping highlights progression of research over the years in terms of numbers of publications, diversity in terms of types of publications, and publication venues as sources of research emergence.

\subsection{Analyzing frequency and the types of the published research}
\label{sec:freq-types-research}
\begin{figure}
  \centering
    \includegraphics[width=0.5\textwidth]{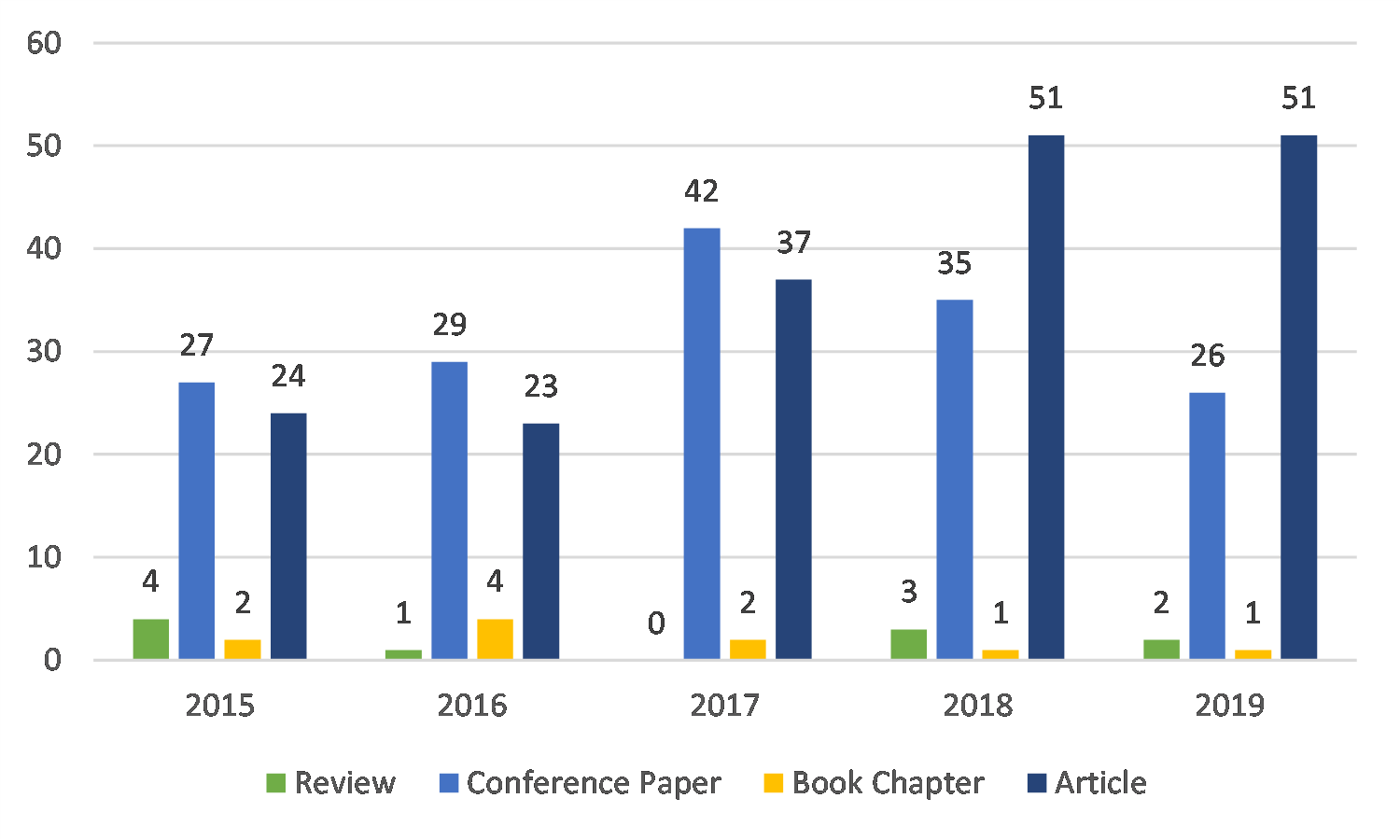}
  \caption{Number of types of research papers along the years.}
  \label{fig:pub-types-years}
\end{figure}

To answer RQ-1, we analysed the types and the frequencies of the published research from years 2015 to 2019 as shown in Figure \ref{fig:pub-types-years}.
As discussed earlier (Section \ref{sec:phase-ii}), through our pilot searches, we could not identify any relevant publication before 2015 on the topic, whereas 2019 (02/10/2019) represents the cut-off point of our literature search as shown in Figure \ref{fig:pub-types-years}.
For fine-grained analysis and interpretation of Figure \ref{fig:pub-types-years}, we provide complementary information in terms of:
(a) types of research publications in Figure \ref{fig:research-type}, 
(b) types of research contributions in Figure \ref{fig:contribution-type}, 
(c) number of publications on family of security and privacy controls in Figure \ref{fig:secpri-families},
(d) number of publications on common types of m/uHealth applications in Figure \ref{fig:application-type}, and
(e) number of publications on technologies being used to enable m/uHealth applications in Figure \ref{fig:used-technology}.
All of the above-mentioned quantitative analysis are detailed later in this section to highlight progression and diversity of the published research on secure and private m/u-Health systems.

Figure \ref{fig:pub-types-years} has two main dimensions. 
The first dimension includes the frequency of the publications in terms of the total number of the studies published during the respective years (projected along y-axis).
The second dimension presents the diversity of the published research in terms of different types of publications such as conference proceedings or journal articles during each year (projected along y-axis).
In this second dimension, the papers are classified as Review, Conference Paper, Book Chapter and Article (as provided by Scopus).
The unified view of Figure \ref{fig:pub-types-years} highlights the total number and the types of the published research.
Figure 3 highlights a gradual increase in the frequency of the research publications such that year 2015 (57 publications), 2016 (57 publications), 2017 (85 publications), 2018 (90 publications) and 2019 (80 publications). 
We interpret the increased frequency as an indication of the growth of the publications that corresponds to an increasing interests in the community for the research and development of security and privacy in the context of m/uHealth systems. 
Initially published studies in the year 2015 such as [P16, P51, P52]~\footnote{The notation [$Pn$], $P$ stands for \textbf{P}rimary studies and $n$ represents a number ($n = 1, 2, …, 365$) refers to 365 primary studies that have been selected for review, in this mapping study, presented in \cite{iwaya2020protocol} (Appendix \ref{app:primary-studies-list}). The notation also maintains a distinction between bibliography ('References'), primary studies for this review (Appendix \ref{app:primary-studies-list}, \cite{iwaya2020protocol}), and the \textbf{E}valuation \textbf{R}esearch ($ER$) (Appendix \ref{app:evaluation-research-studies}, \cite{iwaya2020protocol}).} primarily focused on security, mobile apps, smartphones and sensor networks in the context of m/uHealth systems.
In contrast, the recently published studies from 2019 such as [P286, P289, P322] addresses privacy (vis-a-vis GDPR),``cyber''-security, blockchain and IoT to enable secure and private m/uHealth systems.
Another observation in Figure \ref{fig:pub-types-years} is about more than doubled journal articles that theoretically represent the detailed presentation  of the research challenges, the proposed solutions and rigorous evaluations. %
For example, the study [P346] published in Journal of Medical Internet Research provides fine-grained presentation of an open-source platform for processing mHealth data using sensors, wearables and mobile devices.

\subsubsection{Types of the research publications}
\begin{figure}
  \centering
    \includegraphics[width=0.5\textwidth]{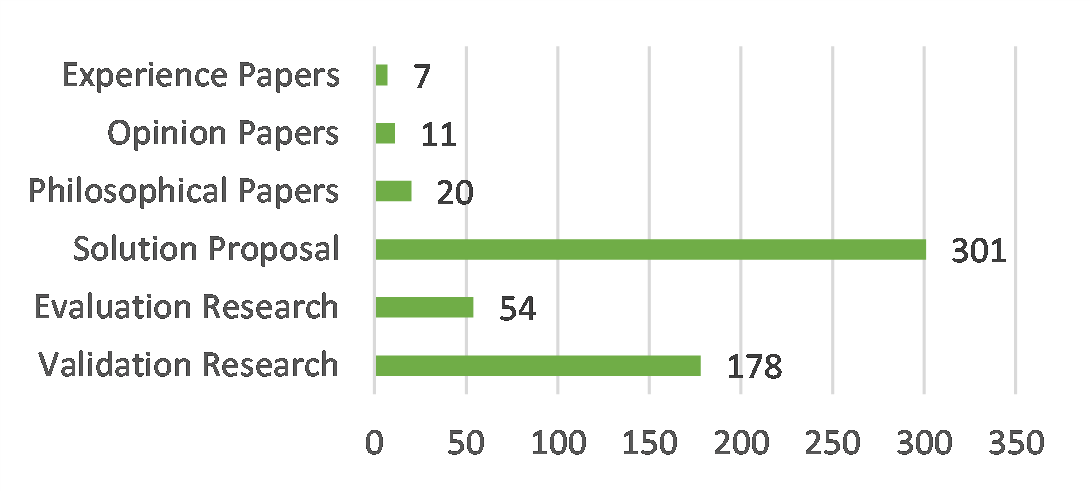}
  \caption{Number of publications based on each research types.}
  \label{fig:research-type}
\end{figure}

Figure \ref{fig:research-type} highlights the types of research publications across 6 different well established categories that have been adopted from \cite{wieringa2006reqs}. 
For example, Figure \ref{fig:research-type} shows that the most number of the publications as Solution Proposals (i.e., 301 studies) and the least number as Experience Papers (i.e., 7 studies only). 
Solution Proposals represent novel ideas that formulate an innovative solution to already established or emerging solutions for security and privacy of m/uHealth system. 
For example, the studies [P58, P135, P337] represent the solution proposals for security and privacy frameworks for ubiquitous IoT systems, sensor-based remote monitoring, and mHealth apps, respectively. 
Further details about each type of the research publications as shown in Figure \ref{fig:research-type} can be found in \cite{wieringa2006reqs}.
The primary intent of Figure \ref{fig:research-type} is to represent the diversity of the published research in the context of the well established categories of the research publication types. 

\subsubsection{Types of the research contributions}
\begin{figure}
  \centering
    \includegraphics[width=0.5\textwidth]{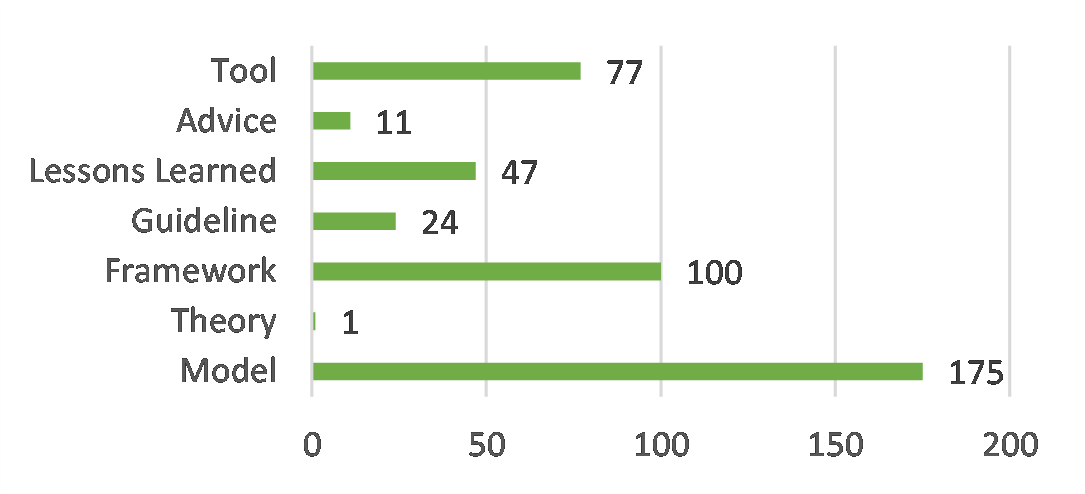}
  \caption{Number of publications based on each contribution types.}
  \label{fig:contribution-type}
\end{figure}

Figure \ref{fig:contribution-type} shows different types of research contributions across 7 distinct categories that have been adopted from \cite{kuhrmann2016software, shaw2003writ}. 
Specifically, Figure \ref{fig:contribution-type} shows that the most types of the research contributions are in the form of Model (i.e., 175 studies) and the least type as Theory (i.e., 01 study). 
Research contributions categorized as Model focus on the representation of the observed reality by concepts.
For example, the studies [P30, P73] that represent a mathematical models in the form of cryptographic schemes for authentication and key agreement.
Further details about each type of the research contribution as shown in Figure \ref{fig:contribution-type} can be found in  \cite{kuhrmann2016software, shaw2003writ}.
The aim of presenting the data in Figure \ref{fig:contribution-type} is to provide a high-level view of different research contributions as the constituent elements of the published literature that vary from tools (enabling automation), to lessons learned (recommendation and guidelines), and frameworks (integrated development environments) to enable or enhance the security and privacy of m/uHealth Systems.

\subsubsection{Number of the publications on the families of the privacy and security controls }
\begin{figure}
  \centering
    \includegraphics[width=0.5\textwidth]{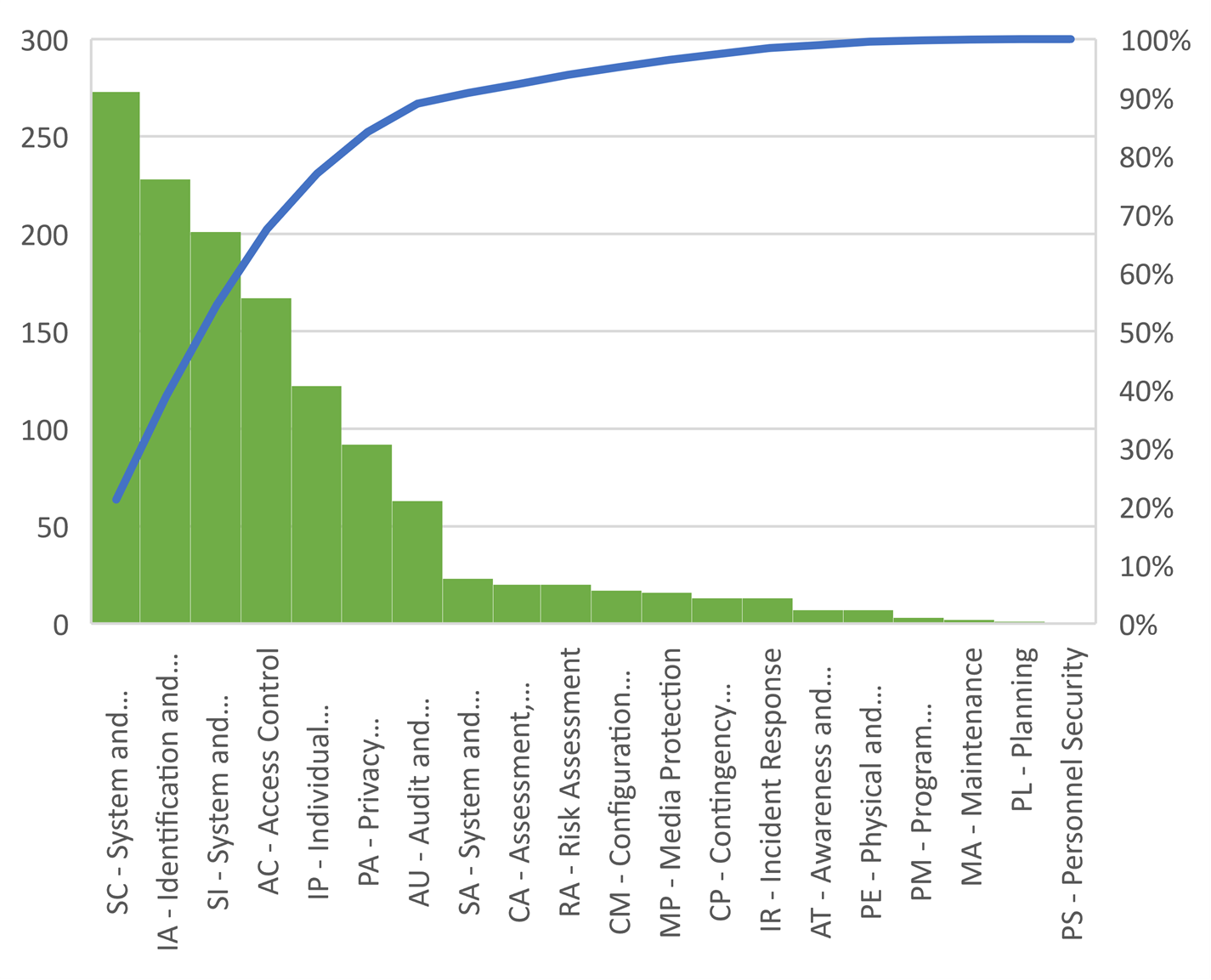}
  \caption{Number of publications based on each family of security and privacy controls.}
  \label{fig:secpri-families}
\end{figure}

Figure \ref{fig:secpri-families} shows studies distributed across a total of twenty security and privacy control families according to NIST-8062 standard \cite{nist8062}.
The vast majority of existing solutions implement controls from the families of system and communication protection (SC, $n=273$) and identification and authentication (IA, $n=228$), which are typical security controls for confidential communication (i.e., encryption), and user or device authentication.
Organisational aspects addressed by families such as planning (PL, $n=1$) and personal security (PS, $n=0$) are under-represented in the existing research.

\subsubsection{Number of publications on common types of the m/uHealth applications}
\begin{figure}
  \centering
    \includegraphics[width=0.5\textwidth]{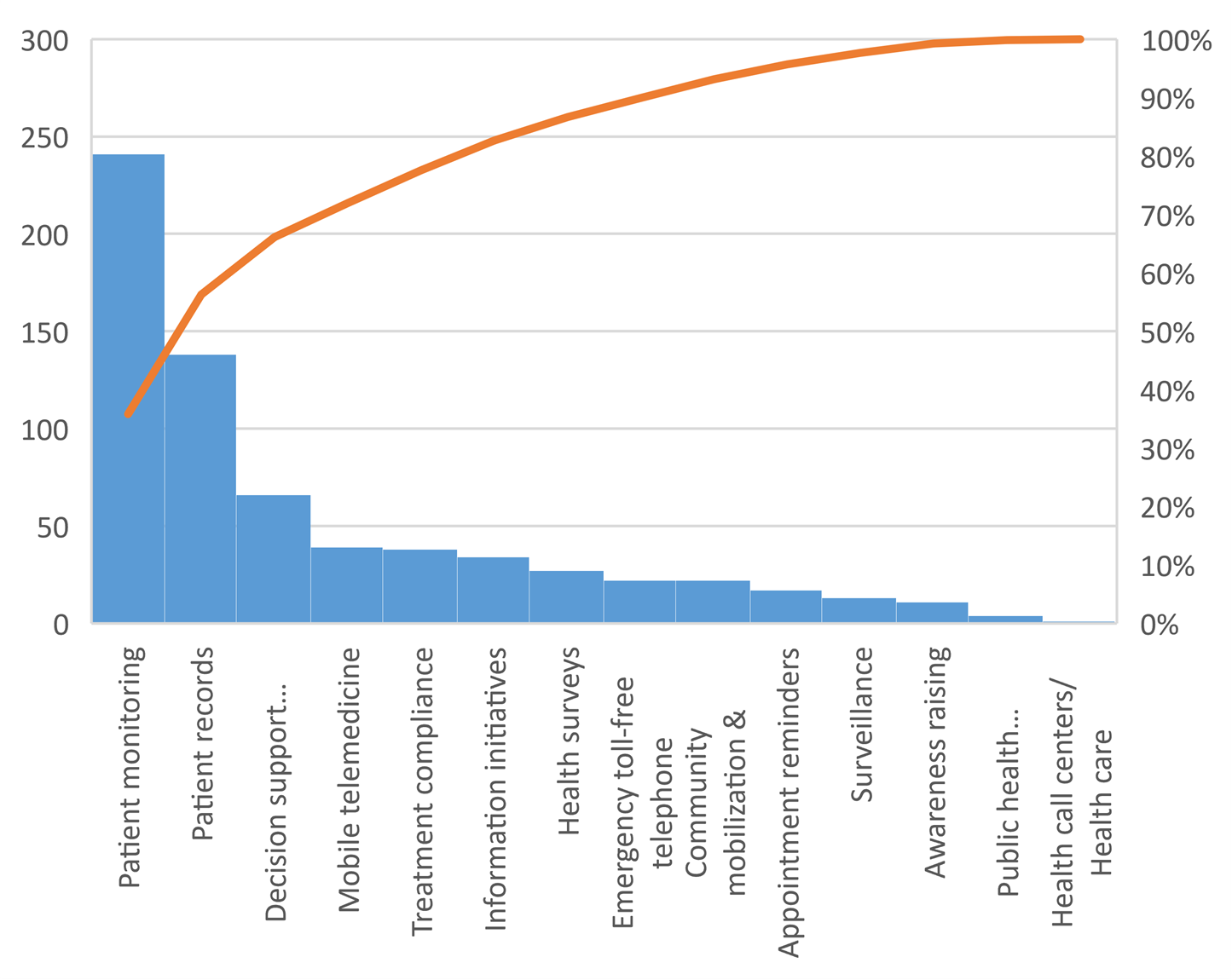}
  \caption{Number of publications based on each application categories.}
  \label{fig:application-type}
\end{figure}

Figure \ref{fig:application-type} presents the different types of health applications that were reported in the reviewed studies.
This classification is based on the World Health Organization (WHO) reports on eHealth and mHealth technologies that have been reported by all the affiliated countries \cite{who2011mhealth}.
The applications for patient remote monitoring ($n=241$) (e.g., use mobile devices or sensor tracking of vital signs) and for patient records ($n=138$) (e.g., personal health records or electronic health records) are still the most traditional mHealth systems.
On the other hand, the mHealth application categories for health call centers and helplines ($n=4$), as well as public health emergencies ($n=4$) have not been the usual targets with respect to security and privacy research.

\subsubsection{Number of the publications on technologies being used to enable m/uHealth applications}
\begin{figure}
  \centering
    \includegraphics[width=0.5\textwidth]{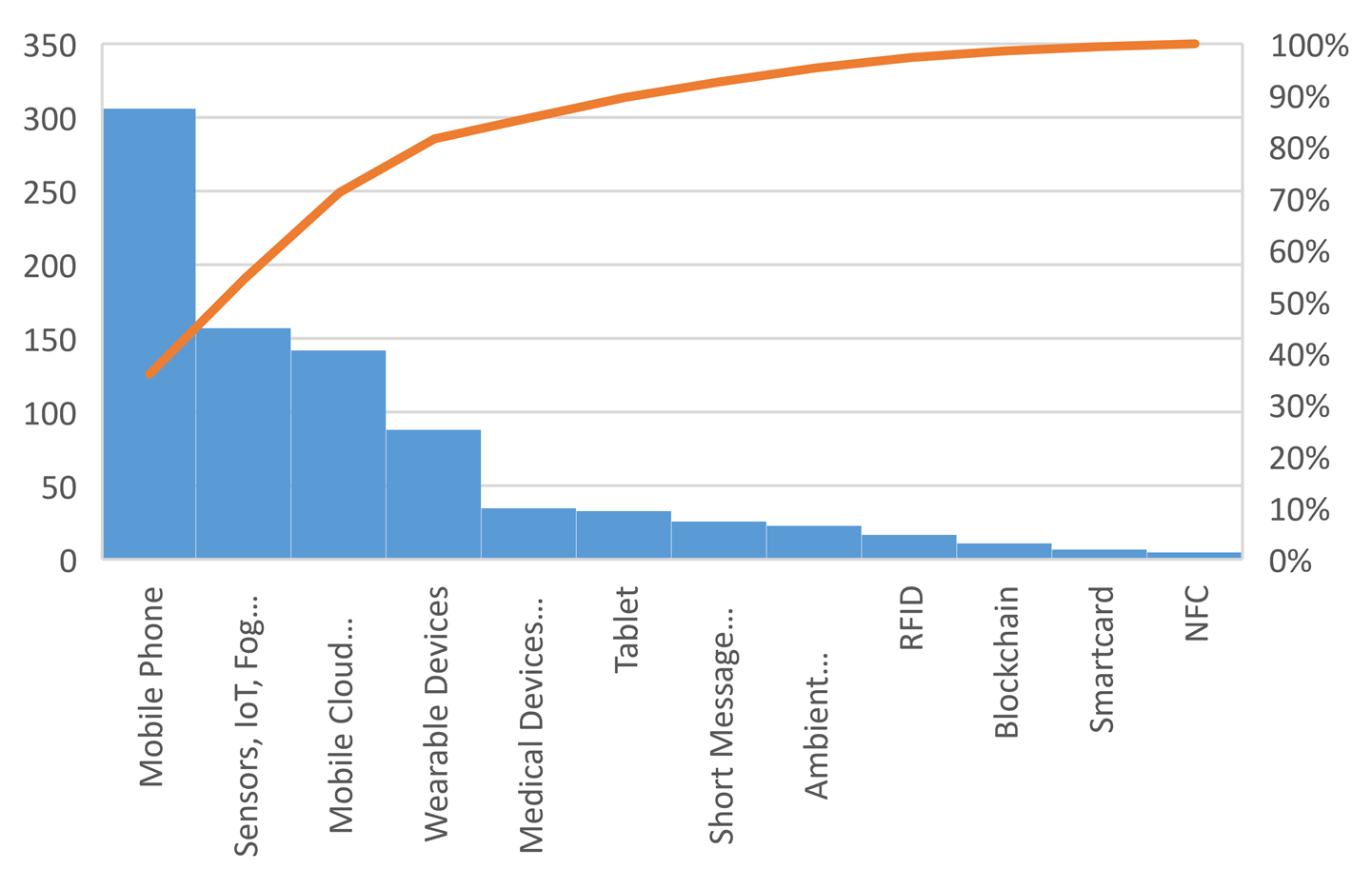}
  \caption{Number of publications based on used technologies.}
  \label{fig:used-technology}
\end{figure}

Figure \ref{fig:used-technology} shows the types of used technologies that enable various m/uHealth systems.
This classification was created by the authors using the keywording method, and thus forming groups of closely related keywords.
Not surprisingly, the use of mobile phones (or smatphones) is by far the most used type of technology ($n=306$), followed by sensors and IoT devices ($n=157$).
Technologies such as RFID ($n=17$), Blockchain ($n=11$), Smartcard ($n=7$) and NFC ($n=5$) have not yet been extensively addressed.

\subsection{Prominent venues of the published research}
\label{sec:prominent-venues}
We now answer RQ-2 that is aimed at identifying and discussing the venues of the publications. 
Table \ref{tab:pub-venues} provides a structured catalogue to provide all the relevant information about the prominent venues of the reviewed publications. 
The prominent venues refer to any journal or conference proceedings etc. with a specific minimum number of publications (i.e., at least 06 published studies).
The results from our pool of the selected publications indicate that an increasing number of papers are being published every year (see Figure \ref{fig:pub-types-years}).
A vast majority of the publications fall into the categories of journal articles ($n=197$) and conference papers ($n=164$).
There are dozens of publications venues where the reviewed papers have been published. That means the published research on security and privacy of m/uHealth is sparsely distributed in different journals, conference proceedings and books.
There are nine venues that cover nearly 20\% of the total number of the reviewed papers, as shown in Table \ref{tab:pub-venues}.
All other venues have the lower number of publication, ranging from 1 to 4 papers only.

\begin{table*}[h!]
\caption{Top 9 publication venues. Number of publications (\#) and total percentages (\%) for journals and conferences.}
\label{tab:pub-venues}
\begin{tabular}{|p{0.3\textwidth}|c|c|c|c|p{0.065\textwidth}|p{0.25\textwidth}|p{0.07\textwidth}|}\hline
\textbf{Source Title and Publisher} & \textbf{\#} & \textbf{\%} & \textbf{Venue} & \textbf{Rank} & \textbf{Acum.  Citations} & \textbf{Ref. to Studies IDs} & \textbf{Publ. Country} \\ \hline\hline
IEEE Access (IEEE) & 15 & 3.93 & Journal & Q1 & 334 & P114, P133, P166, P216, P226, P240, P256, P260, P279, P283, P324, P327, P328, P339, P359 & USA \\
Lecture Notes in Computer Science (Springer) & 12 & 3.14 & Conf. & Q2 & 40 & P39, P67, P74, P82, P87, P128, P232, P262, P277, P314, P329, P342 & Germany \\
Sensors (MDPI) & 9 & 2.36 & Journal & Q2 & 105 & P19, P134, P158, P165, P176, P199, P298, P302, P319 & Switzerland \\
Journal of Medical Internet Research (JMIR) & 9 & 2.36 & Journal & Q1 & 97 & P50, P209, P221, P281, P290, P331, P333, P346, P362 & Canada \\
Journal of Medical Systems (Springer) & 8 & 2.09 & Journal & Q2 & 320 & P16, P32, P78, P91, P110, P205, P253, P351 & USA \\
JMIR mHealth and uHealth (JMIR) & 8 & 2.09 & Journal & -- & 59 & P208, P229, P300, P311, P320, P340, P349, P365 & -- \\
Future Generation Computer Systems (Elsevier) & 7 & 1.83 & Journal & Q1 & 504 & P215, P242, P249, P261, P282, P344, P355 & Netherlands \\
Studies in Health Technology and Informatics (IOS Press) & 6 & 1.57 & Conf. & Q3 & 114 & P46, P54, P89, P129, P186, P312 & Netherlands \\
Advances in Intelligent Systems and Computing (Springer) & 6 & 1.57 & Conf. & Q3 & 10 & P126, P224, P246, P264, P297, P316 & Germany \\ \hline
\end{tabular}
\end{table*}

Table \ref{tab:pub-venues} also shows that the venues are mostly journals positioned in the top quartiles (i.e., Q1 and Q2) according the the Scientific Journal Rankings (JCR~\footnote{\url{https://www.scimagojr.com/journalrank.php}}).
High accumulated citations are particularly presented on journals, namely the Future Generation Computer Systems, the IEEE Access and the Journal of Medical Systems.
Comparatively, journal venues have a much higher accumulated number of citations when compared to the conference venues.
Publishers are concentrated in Europe (i.e., Germany, Switzerland and the Netherlands) and North America (i.e., USA and Canada).
Table \ref{tab:pub-venues} shows a diversity of publication venues that include but not limited to computer science and systems, embedded systems, medical and intelligent systems. 
For example, the publication venue named Future Generation Computer Systems (FGCS) categorized as computer science and systems journal has a total of 7 publications [P215, P242, P249, P261, P282, P344, P355] that specifically focus on security of advanced technologies, such as mHealth social networks, cyber-physical networks, edge computing, IoT and big data systems.

\section{Research mapping based on the existing solutions for the security and privacy of m/uHealth systems}
\label{sec:results-part2}
In this section, we answer RQ-3 driven by the mapping of the existing research as illustrated in Figure \ref{fig:mapping-1} and Figure \ref{fig:mapping-2}, corresponding to our second-stage analysis and documentation (as in Figure \ref{fig:methodology}). 
We provide the research mapping that identifies the state-of-the-art in terms of the existing solutions and their prime contribution(s) to support security and privacy of m/uHealth systems (Section \ref{sec:mapping-existing-solutions}). 

\subsection{Mapping of the existing solution for the secure and private m/uHealth}
\label{sec:mapping-existing-solutions}
This section answers RQ-4 to present a systematic map of the existing solutions. 
First we present the mapping between the types of the research contributions and the types of the security and privacy control families (Section \ref{sec:mapping-contribution-secpri}). 
We then present the mapping between the types of the applications and the types of the technologies being used to support the security and privacy of m/u-Health systems (Section \ref{sec:mapping-applications-technologies})
\subsubsection{Mapping between the Research Contributions and the Security and Privacy Control Families}
\label{sec:mapping-contribution-secpri}
\begin{sidewaysfigure*}
  \centering
    \includegraphics[width=1.0\textwidth]{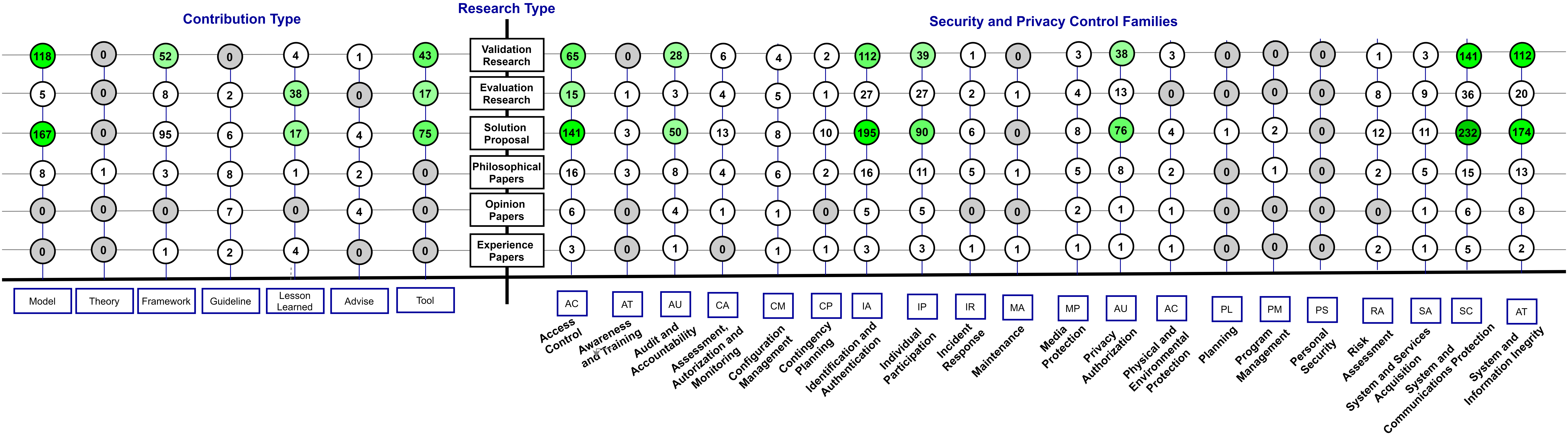}
  \caption{Mapping results positioning the \emph{research type} with \emph{contribution type} and \emph{security and privacy control families}.}
  \label{fig:mapping-1}
\end{sidewaysfigure*}

The first systematic map produced in this SMS is shown in Figure \ref{fig:mapping-1}. 
The systematic map works as a matrix (based on x/y-axis) that unifies (a) three distinct facets to their (b) corresponding evidence as detailed below.

\textit{Facets of systematic mapping} -- According to Figure 9, these are presented as:
\begin{itemize}
    \item Research type facet is adopted from Figure \ref{fig:research-type} that is central to map research contributions and control families, drawn on y-axis of Figure \ref{fig:mapping-1}.
    \item Contribution type facet presented on x-axis at left of Figure \ref{fig:mapping-1} draws different types of research contributions adopted from Figure \ref{fig:contribution-type}.
    \item Security and Privacy control families facet presented on x-axis at the right of Figure \ref{fig:mapping-1} draws different types of security and privacy control families from Figure \ref{fig:secpri-families}.
\end{itemize}

\begin{sidewaysfigure*}
  \centering
    \includegraphics[width=1.0\textwidth]{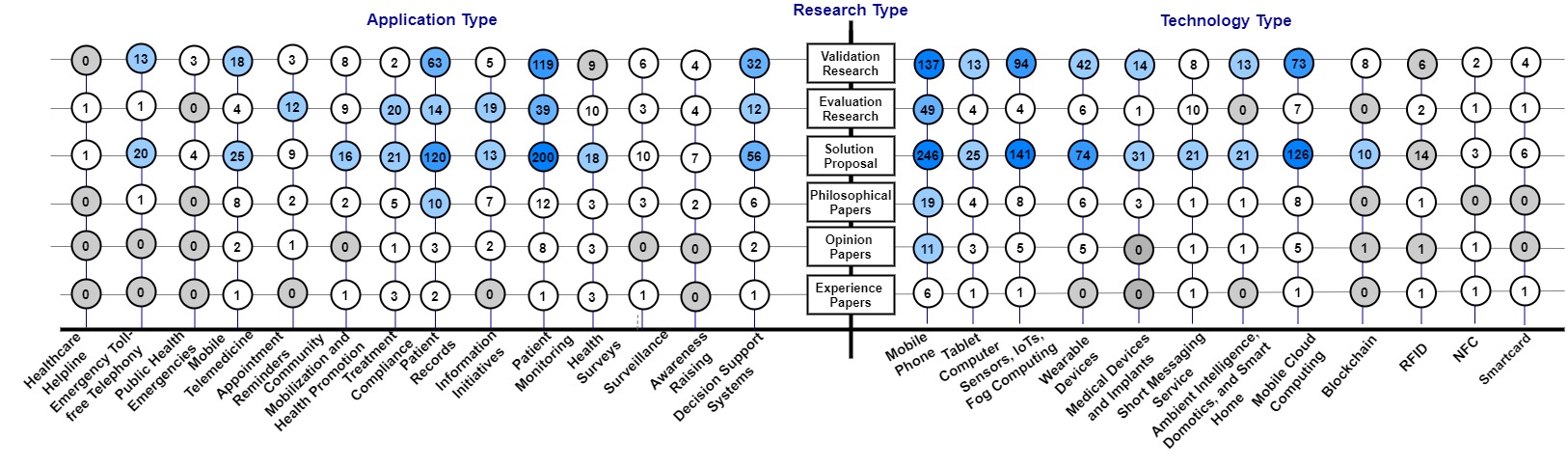}
  \caption{Mapping results positioning the \emph{research type} with \emph{application types} and \emph{used technology}.}
  \label{fig:mapping-2}
\end{sidewaysfigure*}

\textit{Evidence for systematic mapping}  -- The published evidence for the systematic mapping refers to the reviewed studies (Appendix \ref{app:primary-studies-list} \cite{iwaya2020protocol}) that are relevant to a particular map that are presented as bubbles plotted in Figure \ref{fig:mapping-1}.
\begin{itemize}
    \item \textbf{Evidence for Contributions} is presented on y-axis at the left of Figure \ref{fig:mapping-1} that provides a mapping between the research types and its particular contributions. For example, based on this mapping we can interpret that there are a total of 42 published papers (e.g., [P1, P22, P28]) that focus on the validation of the tools developed to enable or enhance the security and privacy of m/uHealth systems. For example, the study reported in [P22] presents a mHealth systems for securely transmitting personal bio data. To validate the proposed tool, the authors implemented it and carried a controlled experiment by spoofing the communication channel to demonstrate confidentiality. Alternatively, we can view that there are no opinion or philosophical papers about tool support for the security and privacy of m/uHealth systems.
    \item \textbf{Evidence for security and privacy control families} is presented on y-axis at the right of Figure \ref{fig:mapping-1} that provides a mapping between the research types and the security and privacy control families. For example, this mapping can help us to interpret that in our mapping study a total of 141 solutions have been proposed (e.g. [P44, P51, P56]) that exploit Access Control as a mechanism to enable or enhance the security and privacy of m/uHealth. For example, a study [P51] presents the solutions as a security framework for mHealth systems that implements a role-based access control mechanisms to achieve authorised access to health data. Based on the mapping, we can also see that there is no validation of Awareness and Training to support security and privacy.
\end{itemize}

Based on Figure \ref{fig:mapping-1}, there can be multiple and diverse interpretations about the existing research. 
An exhaustive detail about each possible types of interpretation is not possible. 
However, we provide some examples and guidelines below that can help a reader to identify and interpret the available information as per their needs. 
These examples include:
\begin{itemize}
    \item \textit{What is the most and least focused security and privacy control families in published literature?} In order to identify the most focused (top 3 by number of publications) security and privacy control families, we have identified System and Communication Protection (273 studies), Identification and Authentication (228 studies) along with System and Information Integrity (201 studies) as the most researched ones. On the contrary, the least focused families include Personal Security (0 studies), Planning (1 studies), and Program Management (2 studies) as least focused areas.
    \item \textit{What types of contributions gain the most or the least attention in the recent years?} The results show that a vast majority of the contributions are in the forms of Models (175 studies), Frameworks (100 studies) along with Tools (77 studies). However, a significantly less expressive number of studies offer the contributions in terms of Theories (1 study), Advice (11 studies) and Guidelines (24 studies) on engineering and development of the secure and private m/uHealth systems.
    \item \textit{What types of research has the most and the least number of publications?} As usual in the area of computer science, most of the research is categorised as Solution Proposal (301 studies), followed by Validation Research (178 studies) and Evaluation Research (54 studies). For example, the studies [P279, P306] as validation research demonstrate that is feasible to perform privacy-preserving search/querying over mHealth data, even though further studies are warranted to evaluate of the proposed solutions in practice.
\end{itemize}

\subsubsection{Mapping between the types of applications and the types of technologies}
\label{sec:mapping-applications-technologies}
The second systematic map produced in this SMS is shown in Figure \ref{fig:mapping-2}.

\textit{Facets of systematic mapping} -- According to Figure \ref{fig:mapping-2}, these are presented as:
\begin{itemize}
    \item \textbf{Research type facet}, as aforementioned, is drawn on y-axis of Figure \ref{fig:mapping-1}, central to map application types and used technologies.
    \item \textbf{Application type facet} presented on the x-axis at left of Figure \ref{fig:mapping-2} draws different types of m/uHealth applications adopted from Figure \ref{fig:application-type}.
    \item \textbf{Used technology facet} presented on x-axis at the right of Figure \ref{fig:mapping-2} draws different types of the used technologies from Figure \ref{fig:used-technology}.
\end{itemize}

\textit{Evidence for systematic mapping} -- The published evidence for systematic mapping that are relevant to this map are presented as bubbles plotted in Figure \ref{fig:mapping-2}.
\begin{itemize}
    \item \textbf{Evidence for Applications} is presented on the y-axis at the left of Figure \ref{fig:mapping-2}, providing a mapping between the research types and its application categories. For instance, it is possible to interpret that a total 200 studies (e.g., [P4, P13, P14]) are solution proposals for the patient monitoring systems. For example, the paper [P13] provides a solution in the form of a mHealth patient monitoring emergency alert system.
    \item \textbf{Evidence for the used technologies} is presented on the y-axis at the right of Figure \ref{fig:mapping-2} that provides a mapping between the research types and the used technologies. This mapping shows for example that 47 studies (e.g., [P23, P62]) that used mobile phones as their enabling technologies have also been evaluated. The work reported in [P62] is an example of a privacy-aware telemonitoring systems for patient with chronic heart failure that has been evaluated in practice.
\end{itemize}

\section{Evaluation studies on the security and privacy of m/uHealth Systems}
\label{sec:results-part3}
In this section we answer RQ-4 by presenting the existing evaluation studies on security and privacy of m/uHealth systems. 
The answer to the above mentioned question help us to investigate the evaluation research (i.e., Research Type facet in Figure \ref{fig:mapping-1} and Figure \ref{fig:mapping-2}).
Investigating the studies on the evaluation research pinpoints the existing or innovative methods and techniques being adopted or developed to validate the proposed solution (presented in Section \ref{sec:results-part2}).
Moreover, an explicit discussion on the solution evaluation and the evaluation studies streamline the rigor of the published research and its validation.

All studies in the Evaluation Research facet intersect the topics of (1) security, (2) privacy and (3) m/uHealth.
However, these studies usually put more emphasis on one of the topics.
For instance, a study that reports a mHealth intervention mainly emphasises the health outcomes (e.g., use of a weight-loss app [ER02]).
Although the authors also claim to have the security and privacy controls implemented, these controls are not the focus of the study or of its evaluation.
A summary of the Evaluation Research in respect to their main emphasis is shown in Table \ref{tab:eval-res-emph}.
As per the methodology, in  this second-stage analysis, we are especially interested in the evaluation process in terms of the security and privacy technologies (as per Table \ref{tab:eval-res-emph}, the studies with the emphasis on 'privacy', 'security', and 'security and privacy', i.e., $n=13+12+3=28$, ) and to the extent that they have been evaluated in the current state of the research.
That is not to say that the evaluation process regarding healthcare and health outcomes is of less importance, rather there has been extensive research on the evaluation and maturity of m/uHealth systems already \cite{beratarrechea2014impact, marcolino2018slr}, as opposed to their security and privacy technologies.

\begin{table}[h!]
\caption{Distribution of the papers with respect to the main emphases.}
\label{tab:eval-res-emph}
\footnotesize
\centering
\begin{tabular}{|p{0.075\textwidth}|c|c|p{0.25\textwidth}|}\hline
\textbf{Emphasis} & \textbf{\#} & \textbf{\%} & \textbf{References}\\ \hline\hline
Healthcare & 24 & 46\% & [ER02], [ER03], [ER05], [ER07], [ER08], [ER09], [ER12], [ER13], [ER21], [ER22], [ER24], [ER25], [ER32], [ER35], [ER38], [ER40], [ER41], [ER43], [ER45], [ER46], [ER49], [ER50], [ER51], [ER52] \\
\textbf{Privacy} & \textbf{13} & \textbf{25\%} & [ER04], [ER06], [ER10], [ER15], [ER19], [ER29], [ER30], [ER31], [ER34], [ER37], [ER39], [ER42], [ER47] \\
\textbf{Security} & \textbf{12} & \textbf{23\%} & [ER11], [ER14], [ER16], [ER17], [ER20], [ER23], [ER26], [ER27], [ER28], [ER33], [ER44], [ER48] \\
\textbf{Security and Privacy} & \textbf{3} & \textbf{6\%} & [ER01], [ER18], [ER36] \\ \hline
Total & 52 & 100\% & -- \\ \hline
\multicolumn{4}{p{0.45\textwidth}}{Notes: Although the studies P24 and P90 are classified as Evaluation Research, their evaluation component is equivalent to the presented in other works of the same authors, i.e., ER20 and ER29 respectively.}
\end{tabular}
\end{table}

In the remainder of this section, first, some general comments are provided on the very few studies that put emphasis in healthcare and that at the same time show a significant amount of evidence of evaluation of its security and privacy components.
Next, an in-depth presentation of the themes is provided, configuring the thematic analysis in five broad groups: 
(1) security and privacy evaluation strategies (Tables \ref{tab:sec-pri-eval-themes} and \ref{tab:sec-pri-eval-themes-2}); 
(2) identified problems in industry (Table \ref{tab:ind-prob-themes}); 
(3) solutions as security and privacy models and frameworks (Table \ref{tab:mod-fram-themes}); 
(4) solutions as security and privacy assessment methodologies (Table \ref{tab:sec-pri-sol-themes}); and, 
(5) solutions as security and privacy systems and applications (Table \ref{tab:sys-apps-themes}).
Lastly, a detailed score card for the quality assessment of the studies with the emphasis on security and privacy is provided in Table \ref{tab:qual-eval}.

\subsection{Overview of the evaluation research}
As per the SMS results, it is often the case that authors are more focused on evaluating health outcomes for a given solution instead of evaluating ``non-functional requirements'' related to security and privacy (see Table \ref{tab:eval-res-emph}).
Other studies are more focused on the performance, feasibility, usability and acceptability of systems. 
For such studies, security and privacy features do exist in the systems, but they are not targeted in the evaluation process.
However, a few exceptions were found in which the studies actually do significant evaluation of the implemented security and privacy technologies, even though the main emphasis is on healthcare.

This study has identified that some studies describe the overall issues, such as lack of compliance with privacy laws, insufficient/lacking privacy policies and app vulnerabilities in their studies (i.e., [ER02, ER07, ER24, ER43]).
A few studies also go as far as proposing or using the methods for privacy policy assessments (i.e., [ER07, ER24]), and privacy models (i.e., [ER05]).
Sometimes, the privacy analysis is part of a broader system specification assessment (i.e., [ER09]) or app assessment criteria (i.e., [ER24]).
Other studies address the security related issues to a greater extent, by means of performing a security analysis (i.e., [ER43]) or implementing a security framework (i.e., [ER41]).
Given that, these studies are further discussed along with the thematic analysis, together with the studies that emphasise security and privacy.

As a general trend, it is also worth noting that half of the evaluation studies ($n=26$, 50\%)  performed a \emph{third-party evaluation} of the existing solutions in the market, e.g., mHealth apps, privacy policies, mobile devices and wearables, communication protocols or standards.
This is very important because many researchers are acting as the third-party evaluator, performing privacy and security analyses and providing an unbiased evaluation of the existing solutions.
As a result, they help identify problems in the industry as well as to provide recommendations to m/uHealth developers and practitioners.

In the following subsections, the studies are categorised in themes, as explained in Section \ref{sec:methodology}.
We have closely clustered the themes in the provided tables to facilitate the analysis.
A sentence that captures an emblematic example is provided for each theme as well as the studies that fall into it.

\subsection{Evaluation Strategies for Security and Privacy}
Many strategies have been used to evaluate the security- and privacy-related models and solutions in the selected studies.
This first group of the themes summarises the evaluation strategies evidenced in the selected studies (see Tables \ref{tab:sec-pri-eval-themes} and \ref{tab:sec-pri-eval-themes-2}), making it the most important group in this thematic analysis.
These evaluation strategies consist of the explicit use of the methods or methodologies for the evaluation purposes.
Some studies propose a more general assessment of apps, represented in the themes \emph{App Critical Content Analysis} and \emph{App General Assessments}.
These studies assess the quality of apps, considering privacy just as one of the sub-components of the assessment (i.e., [ER37, ER24, ER42]).
Nonetheless, in the theme \emph{App Privacy Assessment}, the studies go further, extensively assessing privacy, checking for app's required permissions, consent strategies, privacy policies, transparency and purpose specification (i.e., [ER19, ER29, ER36]).

\begin{table}[h!]
\caption{Distribution of the studies across themes on security and privacy evaluation strategies.}
\label{tab:sec-pri-eval-themes}
\footnotesize
\centering
\begin{tabular}{|p{0.075\textwidth}|p{0.25\textwidth}|c|p{0.05\textwidth}|}\hline
\textbf{Theme} & \textbf{Emblematic Example} & \textbf{\#} & \textbf{Studies}\\ \hline\hline
App Critical Content Analysis & \emph{``We conducted a critical content analysis of promotional (advertising) materials for prominent mental health apps [...]''} [ER37] & 1 & [ER37] \\ 
App General Assessments & \emph{``[...] this paper identifies a set of risk and safety features for evaluating mHealth apps and uses those features to conduct a comparative analysis of the 20 most popular mHealth apps.''} [ER42] & 2 & [ER24] [ER42] \\ \hline
App Privacy Assessment & \emph{``[...] we conducted an analysis of 64 popular self-tracking services to determine the extent to which the services satisfy various dimensions of privacy''} [ER19] & 2 & [ER19] [ER29] [ER36] \\ \hline
App Security Analysis & \emph{``[...] we propose a testing method for Android mHealth apps which is designed using a threat analysis, considering possible attack scenarios and vulnerabilities''} [ER20] & 4 & [ER01] [ER20] [ER36] [ER43] \\
App Security Audit & \emph{``The main objective of this paper is to carry out an audit of security of an mHealth Android application''} [ER11] & 1 & [ER11] \\
App Vulnerability Scanning & \emph{``[...] the objective of the research was to conduct a vulnerability study on mhealth CDM [chronic disease management] apps''} [ER27] & 1 & [ER27] \\ \hline
Formal Rigorous Analysis & \emph{``We prove [mathematically] that the proposed approach achieves differential privacy.''} [ER15] & 1 & [ER15] \\
Formal Security Analysis & \emph{``In this paper, we scrutinize the security structure of the IEEE 802.15.6-2012 standard and perform a security analysis on the cryptographic protocols in the standard.''} [ER48] & 1 & [ER48] \\ \hline
\end{tabular}
\end{table}

\begin{table}[h!]
\caption{(Continuation) Distribution of the studies across themes on the security and privacy evaluation strategies.}
\label{tab:sec-pri-eval-themes-2}
\footnotesize
\centering
\begin{tabular}{|p{0.075\textwidth}|p{0.2\textwidth}|c|p{0.1\textwidth}|}\hline
\textbf{Theme} & \textbf{Emblematic Example} & \textbf{\#} & \textbf{Studies}\\ \hline\hline
Privacy Policy Assessment & \emph{``[...] we analyze the privacy policy of a sample of mHealth apps for breast cancer patients, developing a scale to check if GDPR is complied.''} [ER06] & 11 & [ER01] [ER04] [ER06] [ER07] [ER10] [ER20] [ER30] [ER31] [ER37] [ER39] [ER47] \\
Privacy Policy Transparency Score & \emph{``We reviewed mobile apps retrieved from iTunes and Google Play [...] and evaluated the transparency of data handling procedures of those apps.''} [ER34] & 1 & [ER34] \\ \hline
Server-Side Security Analysis & \emph{``This study aimed to (1) identify relevant security concerns on the server side of mHealth apps, (2) test a subset of mHealth apps regarding their vulnerability to those concerns, and (3) compare the servers used by mHealth apps with servers used in all domains.''} [ER33] & 3 & [ER20] [ER33] [ER36] \\ \hline
Wearable Security Assessment & \emph{``[...] we expose, via an experimental campaign, a methodology to perform a vulnerability assessment (VA) on wearable devices communicating with a smartphone''} [ER23] & 3 & [ER16] [ER23] [ER26] \\
Wireless Comm. Security & \emph{``[...] this paper extends this and investigates the security of ``data-in-transit'' and examines the two wireless communication technologies, Near Field Communication (NFC) and Bluetooth''} [ER17] & 1 & [ER17] \\ \hline
\end{tabular}
\end{table}

Several studies focus on the security-related themes, organised in four main clusters:
(1) \emph{App Security Analysis} (i.e., [ER01, ER20, ER36, ER43]), \emph{App Security Audit} (i.e., [ER11]) or \emph{App Vulnerability Scanning} (i.e., [ER 27]);
(2) \emph{Formal Rigorous Analysis} and \emph{Formal Security Analysis} (i.e., [ER15, ER48]);
(3) \emph{Server-Side Security Analysis} (i.e., [ER20, ER33, ER36]); and,
(4) \emph{Wearable Security Assessments} (i.e., [ER16, ER23, ER26]) and analysis of \emph{Wireless Communication Security} (i.e., [ER17]).

Authors sometimes create and use their own \emph{ad-hoc} evaluation methodology for security and privacy.
However, as detailed in the next sections, many studies either propose a new methodology for the assessments or adopt an existing approach from the literature.
In either case, they contribute by typically evaluating the existing solutions already available in the market.
Also, standing out among the themes in Table \ref{tab:sec-pri-eval-themes}, many studies carried out \emph{Privacy Policy Assessments} or calculate a \emph{Privacy Policy Transparency Score} for a given number of mHealth apps.

\subsection{Evaluating research in the industrial context}
The industry can benefit from several of the reviewed studies that have evaluated the m/uHealth solutions in the market (see Table \ref{tab:ind-prob-themes}).
As a result of the evaluation process, such studies are able to \emph{Identify App Attacks} (i.e., [ER11]) and \emph{Identify App Vulnerabilities} (i.e., [ER01, ER11, ER20, ER27, ER43]).
Another group of studies evaluated wearables and wireless connectivity to \emph{Identify Flawed Protocols} (i.e., [ER48]) and \emph{Wireless Communication Vulnerabilities} (i.e., [ER48]), as well as \emph{Identify Wearable Vulnerabilities} (i.e., [ER16, ER23, ER26]) and \emph{Remote Attacks to HeartBeat-Based Security} (i.e., [ER44]). 
Most remarkably, several studies ($n=9$) reported that the available apps have shown a \emph{Lack of Privacy Policies} (e.g., non-existent, insufficient, unclear or unreadable).
Among such studies, \emph{Compliance} was a recurring theme ($n=13$), whether with respect to the current privacy laws or the security standards, many m/uHealth solutions were found to be non-compliant.

\begin{table}[h!]
\caption{Distribution of the studies across themes on identified problems in industry.}
\label{tab:ind-prob-themes}
\footnotesize
\centering
\begin{tabular}{|p{0.075\textwidth}|p{0.2\textwidth}|c|p{0.1\textwidth}|}\hline
\textbf{Theme} & \textbf{Emblematic Example} & \textbf{\#} & \textbf{Studies}\\ \hline\hline
Compliance & \emph{``Of the twenty-five applications, only one was completely transparent in how it is HIPAA compliant, and only two explicitly mention HIPAA in their terms and agreements in the application.''} [ER01] & 13 & [ER01] [ER02] [ER04] [ER06] [ER19] [ER24] [ER29] [ER31] [ER36] [ER37] [ER42] [ER43] [ER47] \\ \hline
Identify App Attacks & \emph{``As solutions, we propose [...] protection against decompilation, against code analysis, and against modified applications.''} [ER11] & 1 & [ER11] \\
Identify App Vulnerabilities & \emph{``The application security and privacy vulnerabilities results of the analyzed applications was subsequently shared with their development teams.''} [ER01] & 5 & [ER01] [ER11] [ER20] [ER27] [ER43] \\ \hline
Identify Flawed Protocols & \emph{``We show that some protocols have subtle security problems and are vulnerable to different attacks.''} [ER48] & 1 & [ER48] \\
Identify Wearable Vulnerabilities & \emph{``In empirical tests three out of four devices showed a total lack of effective security measures.''} [ER16] & 3 & [ER16] [ER23] [ER26] \\
Remote Attack to HeartBeat-Based Security & \emph{``[...] we discuss the threat of remote attacks on HBBS and highlight that only identifiers which are generated from electrical cardiac recordings are safe from such attacks''} [ER44] & 1 & [ER44] \\
Wireless Comm. Vulnerabilities & \emph{``We show that some protocols have subtle security problems and are vulnerable to different attacks.''} [ER48] & 1 & [ER48] \\ \hline
Lack of Privacy Policies & \emph{``The assessment of data privacy and security showed that the privacy policy was not available for 29.2\% of the apps.
''} [ER07] & 9 & [ER07] [ER24] [ER30] [ER31] [ER34] [ER37] [ER39] [ER42] [ER47] \\ \hline
\end{tabular}
\end{table}

\subsection{Evaluation research -- solutions as models and frameworks}
Despite the evaluation processes \emph{per se}, we also provide a panorama of the security and privacy \emph{solutions} that have been proposed and evaluated in the context of m/uHealth.
Solutions essentially fall into three groups: 
(1) security and privacy models and frameworks (theoretical); 
(2) security and privacy assessment methods (risk-driven analysis); and, 
(3) security and privacy systems and applications (technologies).

Models and frameworks (see Table \ref{tab:mod-fram-themes}) depart from a theoretical solution for a problem, that are later implemented and evaluated in practice.
Some proposed \emph{Security frameworks}, using a wide array of cryptographic mechanisms for achieving CIA (i.e., [ER12, ER41]).
A couple of the reviewed papers have developed specific cryptographic mechanisms for \emph{Key Distribution Schemes} (i.e., [ER18]) or \emph{Privacy Data Release} (i.e., [ER15]).
There is also a study describing a \emph{Privacy model} for mHealth apps (i.e., [ER05]) and another defining a \emph{Security testing environment} to analyze the app's communication and data handling (i.e., [ER01]).

\begin{table}[h!]
\caption{Distribution of the studies across themes on models and frameworks solutions.}
\label{tab:mod-fram-themes}
\footnotesize
\centering
\begin{tabular}{|p{0.075\textwidth}|p{0.25\textwidth}|c|p{0.05\textwidth}|}\hline
\textbf{Theme} & \textbf{Emblematic Example} & \textbf{\#} & \textbf{Studies}\\ \hline\hline
Conceptual Security Framework & \emph{``We also describe the conceptual framework adopted to address the hospital security requirements for implementation. ''} [ER12] & 1 & [ER12] \\
Security Framework & \emph{``A data security framework was designed to ensure the security of data, which was stored locally and transmitted over public networks.''} [ER41] & 1 & [ER41] \\ \hline
Key Distribution Scheme & \emph{``A key distribution scheme based on a group send-receive model (GSRM) is proposed for secure data transmission in wireless sensor networks [...]''} [ER18] & 1 & [ER18] \\
Privacy Data Release & \emph{``[...] we propose an efficient and privacy-preserving mHealth data release approach for the statistic data with the objectives to preserve the unique patterns in the original data bins [...]''} [ER15] & 1 & [ER15] \\ \hline
Privacy Model & \emph{``[...] we developed the Privacy Model for Mobile Data Collection Applications (PM-MoDaC) specifically tailored for apps that are related to the collection of mobile data, consisting of nine proposed privacy measures [...]''} [ER05] & 1 & [ER05] \\ \hline
Security Testing Environment & \emph{``Our testing environment included a Lenovo Yoga laptop running Kali Linux Rolling and a Samsung Galaxy S-6 running Android 7.0 (Nougat).''} [ER01] & 1 & [ER01] \\ \hline
\end{tabular}
\end{table}

\subsection{Evaluation research -- solutions as security and privacy assessments methods}
A major part of the solutions, in the evaluation studies, is devoted to various approaches for assessing the security and privacy in m/uHealth systems.
These solutions tend to be risk-driven, proposing new methodologies for identifying threats, calculating risks and selecting controls.
As mentioned, some methodologies take on a broader perspective, yet addressing security and privacy as part of their \emph{App Assessment Criteria} and \emph{System Specification Assessment} (i.e., [ER24, ER42, ER09]).
Only one study proposes an \emph{App Security and Privacy Assessment Methodology} reviewing the aspects of both, security \emph{and} privacy, of various apps (i.e., [ER36]).
Geared toward privacy, some studies specialise in \emph{Information Privacy Risk Scores} and \emph{Privacy Heuristic Evaluation Methods} (i.e., [ER10, ER19]), as well as in \emph{Privacy Policy Assessment Methods} and \emph{Privacy Policy Assessment Scales} (i.e., [ER04, ER34, ER06]).
The last cluster of the studies focuses on security, tailoring the assessments toward \emph{Security Assurance Recommendations}, \emph{Security Check of Android Apps}, \emph{Security Testing Methods} and \emph{Security Vulnerability Assessment Methods} (i.e., [ER27, ER11, ER20, ER23, ER26]).

\begin{table}[h!]
\caption{Distribution of the studies across themes on security and privacy assessment solutions.}
\label{tab:sec-pri-sol-themes}
\footnotesize
\centering
\begin{tabular}{|p{0.075\textwidth}|p{0.25\textwidth}|c|p{0.05\textwidth}|}\hline
\textbf{Theme} & \textbf{Emblematic Example} & \textbf{\#} & \textbf{Studies}\\ \hline\hline
App Assessment Criteria & \emph{``[...] a novel approach to app assessment could identify high-quality and low-risk health apps in the absence of indicators such as National Health Service (NHS) approval.''} [ER24] & 2 & [ER24] [ER42] \\ \hline
App Security and Privacy Assessment Methodology & \emph{``Long term analyses of the life cycle of the reviewed apps and our general data protection regulation compliance auditing procedure are unique features of the present paper.''} [ER36] & 1 & [ER36] \\ \hline
Information Privacy Risk Score & \emph{``We identify six information privacy risk factors by downloading mHealth apps from the iOS and Android app stores and surveying them with respect to their information privacy risks.''} [ER10] & 1 & [ER10] \\
Privacy Heuristic Evaluation Method & \emph{``We did this by introducing a set of heuristics for evaluating privacy characteristics of self-tracking services.''} [ER19] & 1 & [ER19] \\ \hline
Privacy Policy Assessment Method & \emph{``The template used in this study is useful for developers as regards: (1) The evaluation of the existing privacy policies of their apps for further improvements, (2) setting the privacy policy of their apps for the first time, all by complying with the standards regulating its content.''} [ER04] & 2 & [ER04] [ER34] \\
Privacy Policy Assessment Scale & \emph{``Our scale defines a score for every mHealth app based on several GDPR items that must be complied.''} [ER06] & 1 & [ER06] \\ \hline
Security Assurance Recommendation & \emph{``[...] mobile applications related frameworks and guidelines were reviewed to come up with the security assurance recommendations for mhealth CDM apps [...]''} [ER27] & 1 & [ER27] \\
Security Check of Android App & \emph{``Applying to the source code techniques of reverse engineering, we will try to perform an analysis that allows us to carry out the security check of the Android application HeartKeeper. [...] It can be applied to audit security on any other Android application.''} [ER11] & 1 & [ER11] \\
Security Testing Method & \emph{``we propose a testing method for Android mHealth apps which is designed using a threat analysis, considering possible attack scenarios and vulnerabilities specific to the domain''} [ER20] & 2 & [ER20] \\
Security Vulnerability Assessment Method & \emph{``Specific configuration of our evaluation testbed and then covers the methodology used for our testing procedures.''} [ER26] & 2 & [ER23] [ER26] \\ \hline
System Specification Assessment & \emph{``We describe a ``system specification assessment'' of several SMS text messaging applications [...]''} [ER09] & 1 & [ER09] \\ \hline
\end{tabular}
\end{table}

\subsection{Evaluation research -- solutions as security and privacy technologies}
The last group of the themes refers to two particular studies that describe the implementation of the security and privacy controls.
The first one is a \emph{Graphical User Interface for Communicating Privacy Risks} as a transparency-enhancing tool, allowing users to analyse and compare apps in respect to privacy risks (i.e., [ER10]).
Another solution is a \emph{Trust-Based Intrusion Detection Systems for Medical Smartphone Networks} that establish a level of trust among devices used in healthcare environments (i.e., [ER28]).

\begin{table}[h!]
\caption{Distribution of studies across themes on systems and applications solutions.}
\label{tab:sys-apps-themes}
\footnotesize
\centering
\begin{tabular}{|p{0.075\textwidth}|p{0.25\textwidth}|c|p{0.05\textwidth}|}\hline
\textbf{Theme} & \textbf{Emblematic Example} & \textbf{\#} & \textbf{Studies}\\ \hline\hline
GUI for Communicating Privacy Risks (TETs) & \emph{``The information privacy risk score and the graphical user interface are designed to enable users to better comprehend information privacy risks across multiple apps''} [ER10] & 1 & [ER10] \\ \hline
Trust-Based IDS for MSNs & \emph{``In this work, we focus on the detection of malicious devices in MSNs [Medical Smartphone Networks] and design a trust-based intrusion detection approach based on behavioral profiling.''} [ER28] & 1 & [ER28] \\ \hline
\end{tabular}
\end{table}

\subsection{Quality assessment}
Apart from the thematic analysis, we also assessed the quality of the evaluation studies using eleven research dimensions and the quality of the reported evidences (see Table \ref{tab:qual-eval}).
As explained in Section \ref{sec:methodology}, for this quality assessment we followed the approach introduced in \cite{chen2011evaluation}.
Overall, the studies received an average score of $8.57$ (out 11) on the quality assessment.
A vast majority of the evaluation studies offers a sound research paper, with clearly defined aims and contributions.
Main points of concern are around the research design and sampling items, which consequently also negatively affect other factors.
For example, one study [ER04] has a cumulative score of 9.5 (see Table \ref{tab:qual-eval}), that indicates that the study is indeed a research paper (Res), i.e., not only a viewpoint or opinion, with clear aims (Aim) and context (Con), and details its research design (RDs).
In this study, however, the sample (Sam) was rather small, and there is no control group (Ctr), which affects the data collection (DCo) data analysis (DAn).
Lastly, in this study, the authors also explicitly stated the possible biases (Bia), main findings (Fin); and value (Val) to other areas.

\begin{table*}[h!]
\caption{Detailed score card for the quality assessment of the studies in security and privacy.}
\label{tab:qual-eval}
\footnotesize
\centering
\begin{tabular}{|c|c|c|c|c|c|c|c|c|c|c|c|c|}\hline
\textbf{Ref} & \textbf{1 Res} & \textbf{2 Aim} & \textbf{3 Con} & \textbf{4 RDs} & \textbf{5 Sam} & \textbf{6 Ctr} & \textbf{7 DCo} & \textbf{8 DAn} & \textbf{9 Bia} & \textbf{10 Fin} & \textbf{11 Val} & \textbf{Tot}\\ \hline\hline
ER04 & 1 & 1 & 1 & 1 & 0.5 & 0 & 0.5 & 0.5 & 1 & 1 & 1 & 9.5 \\
ER06 & 1 & 1 & 1 & 0.5 & 1 & 0 & 1 & 1 & 1 & 1 & 1 & 9.5 \\
ER10 & 1 & 1 & 1 & 0.5 & 1 & 0 & 1 & 1 & 1 & 1 & 1 & 9.5 \\
ER15 & 1 & 1 & 1 & 1 & 0.5 & 0 & 0.5 & 1 & 0.5 & 0.5 & 0.5 & 7.5 \\
ER19 & 1 & 1 & 1 & 1 & 1 & 0 & 1 & 1 & 1 & 1 & 1 & 10 \\
ER29 & 1 & 1 & 1 & 0.5 & 0.5 & 0 & 1 & 1 & 1 & 1 & 1 & 9 \\
ER30 & 1 & 1 & 1 & 0.5 & 1 & 0 & 1 & 1 & 1 & 1 & 0.5 & 9 \\
ER31 & 1 & 1 & 1 & 0.5 & 0.5 & 0 & 1 & 1 & 1 & 1 & 0.5 & 8.5 \\
ER34 & 1 & 1 & 1 & 1 & 1 & 0 & 1 & 1 & 1 & 1 & 1 & 10 \\
ER37 & 1 & 1 & 1 & 1 & 1 & 0 & 1 & 1 & 1 & 1 & 0.5 & 9.5 \\
ER39 & 1 & 1 & 1 & 1 & 1 & 0 & 1 & 1 & 1 & 1 & 1 & 10 \\
ER42 & 1 & 1 & 1 & 0.5 & 0.5 & 0 & 0.5 & 0.5 & 1 & 0.5 & 0.5 & 7 \\
ER47 & 1 & 1 & 1 & 1 & 1 & 0 & 1 & 1 & 1 & 1 & 1 & 10 \\ \hline
ER11 & 1 & 1 & 1 & 0.5 & 0 & 0 & 0.5 & 0.5 & 0 & 0.5 & 0.5 & 5.5 \\
ER14 & 0.5 & 1 & 1 & 0.5 & 0 & 0 & 0.5 & 0.5 & 0.5 & 0.5 & 0.5 & 5.5 \\
ER16 & 1 & 1 & 1 & 1 & 0.5 & 0 & 0.5 & 0.5 & 1 & 0.5 & 0.5 & 7.5 \\
ER17 & 1 & 1 & 1 & 0.5 & 1 & 0 & 0.5 & 0.5 & 1 & 0.5 & 0.5 & 7.5 \\
ER20 & 1 & 1 & 1 & 1 & 1 & 0 & 1 & 0.5 & 1 & 1 & 1 & 9.5 \\
ER23 & 1 & 1 & 1 & 0.5 & 0.5 & 0 & 1 & 1 & 1 & 0.5 & 0.5 & 8 \\
ER26 & 1 & 1 & 1 & 1 & 0.5 & 0 & 1 & 1 & 1 & 0.5 & 0.5 & 8.5 \\
ER27 & 1 & 1 & 1 & 0.5 & 1 & 0 & 1 & 0.5 & 1 & 0.5 & 0.5 & 8 \\
ER28 & 1 & 1 & 1 & 0.5 & 0.5 & 0 & 0.5 & 0.5 & 0.5 & 0.5 & 0.5 & 6.5 \\
ER33 & 1 & 1 & 1 & 1 & 1 & 0 & 1 & 1 & 1 & 1 & 1 & 10 \\
ER44 & 1 & 1 & 1 & 1 & 1 & 0 & 1 & 1 & 1 & 1 & 1 & 10 \\
ER48 & 1 & 1 & 1 & 1 & 0 & 0 & 0 & 1 & 1 & 1 & 1 & 8 \\ \hline
ER01 & 1 & 0.5 & 1 & 1 & 1 & 0 & 1 & 1 & 1 & 0.5 & 1 & 9 \\
ER18 & 1 & 1 & 1 & 0.5 & 1 & 0 & 0.5 & 0.5 & 0.5 & 1 & 1 & 8 \\
ER36 & 1 & 1 & 1 & 1 & 0.5 & 0 & 1 & 1 & 1 & 1 & 1 & 9.5 \\ \hline \hline
Tot & 27.5 & 27.5 & 28 & 21.5 & 20 & 0 & 22.5 & 23 & 25 & 22.5 & 21.5 & 241 \\ \hline
AVG & 0.98 & 0.98 & 1 & 0.77 & 0.71 & 0 & 0.8 & 0.82 & 0.89 & 0.8 & 0.77 & 8.57 \\ \hline
\multicolumn{13}{p{0.8\textwidth}}{\textbf{Notes:} Research paper (Res); Aims (Aim); Context (Con); Research Design (RDs); Sampling (Sam); Control Group (Ctr); Data Collection (DCo); Data Analysis (DAn); Bias (Bia); Findings (Fin); Value (Val).}
\end{tabular}
\end{table*}

For the studies that achieved a cumulative score of 10 (e.g., [ER19, ER34]), the authors have done an excellent work in detailing all aspects of the reported research.
However, the studies with low cumulative score (e.g., [ER11, ER14]), we found that important explanations were missing, particularly on the research design as well as details on data collection and analysis, which negatively impacted the overall research findings and value.
Besides, it is also worth noting that none of the evaluation studies have reported using a control group in their research.

We found that 15 (out of 28) studies presented a clear and detailed description of the research design and methodology.
Authors of these studies have carefully created or adopted assessments artefacts as part of their methodology, making it clear what is going to be evaluated and how, enabling a high-level of research reproducibility.
Most of the studies evaluated the existing apps, their privacy policies or wearable device, and their authors would typically select a sample of a category of applications for this evaluation.
In 13 of the evaluation studies, the samples were small, which limits the extent of the evaluation of a new method for the security or privacy assessment, and thus also limiting the generalisability of the findings and the overall research value.

Regarding the data collection and analysis, 18 studies on each category provided sufficient description in their methodologies.
Other studies mentioned their data collection and analysis procedures, but the descriptions were too brief (e.g., lacking information about sources, surveys, or interview processes) or did not provide enough explanation and justification.
Most of the studies ($n=24$) defined an evaluation strategy to assess, test or scan security and privacy issues, so that the authors could have fairly evaluate the solutions in an unbiased manner.
Lastly, we considered that the scores for the research findings and value were 17 and 15 respectively, mainly looking if the authors were clearly stating the research findings and the value of the research to the broader community.
However, we found that the findings and the value scores were also negatively influenced by the quality of the research design and other processes.

\section{Review and discussion of the SMS results}
\label{sec:discussion}
After answering all the RQs individually, in this section, we review and discuss the core findings of the mapping study on the security and privacy of m/uHealth systems.
This discussion is aimed at demonstrating the collective impact of the existing research in terms of its strengths and limitations by highlighting  the under-/over-researched areas, perceived trends and potential gaps based on the evidences from the literature. 
In the remainder of this section, we refer to the respective figures (Figure \ref{fig:pub-types-years} to Figure \ref{fig:mapping-2}) as illustrations and tables as structured information (Table \ref{tab:ind-prob-themes} to Table \ref{tab:qual-eval}) that complement the review provided here.

\subsection{Stage I analysis: types, frequency, and venues of the research publications}
Stage I analysis primarily focus on demographics of the published research, specifically in terms of types, frequency and prominent venues of the research publications that are reviewed below. 

\subsubsection{Types and frequency of the published research (RQ-1)}
The results in Figure \ref{fig:pub-types-years} shows a steady increase in the number of publications on the topic.
Over the years since the first published study on the topic in 2015, 90\% of the publications have been published in scientific journals and conference proceedings.
Most of the contributions, as shown in Figure \ref{fig:research-type}, are in the form of solution proposals, usually applying some kind of validation through prototyping, performance analysis or user studies.
However, only about 60/365 (i.e., 16.5\%) publications have reported proper evaluation of the systems, i.e., with real-life deployment and medium to large scale testing with end users, such as [ER40, ER41].
Although validation is the first step before starting any stage evaluation, it is important that researchers take such initiatives further and deploy the proposed solutions in real systems with real users.
Validation of prototypes tend to be of rather limited scope, with simplified assumptions that often do not represent the level of complexity that exists in the environment of m/uHealth technologies, such as [P279, P306].
Validation methods also tend to focus on bench-marking a set of variables for performance analysis (e.g., time consumption, memory requirements or usability), instead of fully engaging with all users, stakeholders, and real problems that need to be solved.
Specifically, the reported solutions tend to compete for performance benchmarks or even argue for higher levels of security and privacy, but the question remains that do they really address an existing problem in people's lives? What is the impact of the improved performance?
Essentially, rigorous evaluation brings to surface the more tangible relevance of the research, so as to avoid creating solutions to artificial problems, or even starting with a solution and only then trying to find the problem to use it.

\subsubsection{Prominent venues of the publications (RQ-2)}
Another important finding relates to the publication venues (in Table \ref{tab:pub-venues}).
The area of m/uHealth has seen much growth during the past decade, but the contributions are rather sparsely distributed in a large number of scientific journals and conferences.
In fact, this SMS does not seem to reveal any major leading venue in the field.
Even the top nine venues listed in Table \ref{tab:pub-venues} only represent 20\% of the mapped studies such as [P114, P134], leaving a large majority, almost 80\%, of the studies are thinly distributed among dozens of other publication venues.
These top venues, however, do include prominent journals with significant impact factors as well as proceedings from well-known conferences.
It is also worth noticing that more than half of these top venues have a rather general research scope and four of them are in the field of health informatics as a broad.
The topic of security and privacy for m/uHealth is actually highly specialised and the existing research frequently falls into sub-tracks of the existing journals and conferences.
For instance, the \textit{Journal of Medical Internet Research} has the sister \textit{Journal of mHealth and uHealth}, which has security and privacy as one of its tracks.

\subsection{Stage II analysis: the existing solutions and validation research}
Stage II analysis present the mapping of the existing solutions and their evaluations for engineering secure and private m/u-Health systems that are reviewed below.

\subsection{Existing solutions of the security and privacy of m/uHealth (RQ-3)}
The systematic map presented in Figure \ref{fig:mapping-1} allows us to observe an interesting phenomena that most publications ($\approx 85\%$) match only six out of the twenty control families.
In these twenty families, NIST's special publication 800-53 provides us with a total of 276 security and privacy controls.
The top six families with the most matches (see Figure \ref{fig:secpri-families}) account for a total of 105 controls (i.e., $SC=41, AC=23, SI=19, IA=12, IP=6, PA=4$).
A closer look at these six families reveals that they cover the most fundamental security and privacy controls, such as confidential communication (encryption) (e.g., [P160, P161]), authentication and key exchange (e.g., [P29, P30]), information integrity (e.g., [P139, P140]), authorisation and access control [P146, P159], individual participation and sharing of personal data (e.g., [P185]), and audit and accountability (e.g., [P182]).
These controls are mostly technical, i.e., the use technology to reduce vulnerabilities and can be installed and configured providing protection automatically.
On the one hand, the research focus on such controls exhibit an emphasis from the community about getting the basics of the security and privacy right, given that their correct implementation can indeed solve a vast majority of threats faced by m/uHealth systems.
On the other hand, less focus is given to organisational and managerial controls (e.g., [P81, P358]).
Such under-represented control families are however critical to the operation of organisations using m/uHealth systems.
There are a few hypotheses that could explain such observation: 
a) these areas are not viewed as critical as the others; 
b) researchers in security and privacy as well as m/uHealth practitioners may lack expertise in such areas; 
c) researchers may have a technical bias, and prefer to limit the scope to specific requirements of m/uHealth systems rather than organisational ones.
Nonetheless, implementing organisational controls such as data governance strategies and establishing security and privacy policies are indeed critical to the successful operation of any organisation, and even more if processing personal health data.
More emphasis on such controls families is in fact warranted, specially for m/uHealth initiatives being deployed in low- and middle-income countries, usually covering highly vulnerable populations.

Future research could focus on the under-represented control families.
Many of these controls families tend to address vulnerabilities related to broader aspects of an organisation (e.g., risk assessment, incident response, awareness and training and program management), instead of purely technical controls that can be implemented in a particular m/uHealth system.

It is important to stress that strategies of security and privacy by design are technical approaches to a social problem \cite{gurses2014priveng}, i.e., the fundamental human right to privacy.
Security mechanisms and privacy-enhancing technologies cannot fix a broken business model.
Organisational change is therefore indispensable for establishing a culture of security and privacy across all departments and divisions in a company.

\subsection{Evaluation studies on the security and privacy of m/uHealth systems (RQ-4)}
As described in Tables \ref{tab:sec-pri-eval-themes} and \ref{tab:sec-pri-eval-themes-2}, most of the studies either evaluate a set of mHealth apps or their privacy policies (e.g., [ER04, ER06, ER34]).
Empirical evidence on evaluation concentrates mainly in mHealth systems instead of uHealth systems (e.g., involving wearables, sensors networks or IoT in general).
The reviewed studies used various approaches to evaluate the existing solutions, which we organised in a number of groups and themes.
The first group of the evaluation strategies (Table \ref{tab:sec-pri-eval-themes}), comprises the themes of \emph{App Critical Content Analysis} and \emph{App General Assessments} with a total of 3 studies [ER37, ER24, ER42].
These studies prescribe a general assessment, geared towards quality and safety, considering privacy to a limited extent, which is not sufficient for security and privacy compliance.

On the other hand, two other groups concentrate on such specialised assessments (Table \ref{tab:sec-pri-eval-themes}), specifically for mHealth apps.
For privacy, the theme of \emph{App Privacy Assessments} containing 3 studies [ER19, ER29, ER36], and for security the group of themes on \emph{App Security Analysis}, \emph{App Security Audit} and \emph{App Vulnerability Scanning} with 6 studies in total [ER01, ER20, ER36, ER43, ER11, ER27].
These studies offer more suitable approaches for evaluating security and privacy of mHealth apps.
Nonetheless, the assessment strategies vary greatly, and the future research could integrate these security/privacy assessments in an unified framework for mHealth apps.
Such integrated frameworks can also include the work on the theme of \emph{Server-Side Security Analysis} ((Table \ref{tab:sec-pri-eval-themes-2})), i.e., jointly considering apps, servers, and third-party servers.
This would be useful as a recommendation system built in line with the existing security and privacy standards and regulations in the health sector.

The privacy policies for mHealth apps were the most commonly evaluated artefacts (Table \ref{tab:sec-pri-eval-themes-2}), consisting of the themes of \emph{Privacy Policy Assessment} with 11 studies [ER01, ER04, ER06, ER07, ER10, ER20, ER30, ER31, ER37, ER39, ER47] and \emph{Privacy Transparency Score} with 1 study [ER34].
These assessments focus on the transparency and openness of mHealth apps, i.e., clearly defining what personal data is collected and processed and for which purposes, in easy and understandable language.
This study has also revealed that there are a diverse set of privacy evaluation approaches that tend to rely on manual inspection as well as automatic measurement of text complexity.
We see an opportunity for further research on comparing and integrating the evaluation approaches.
Making privacy policies more understandable is critical for achieving users' \emph{informed consent} under different  privacy regulations (e.g., EU GDPR).

Lastly, on the evaluation approaches, the themes of \emph{Wearable Security Assessment} and \emph{Wireless Communication Security} with 3 and 1 studies respectively (Table \ref{tab:sec-pri-eval-themes-2}).
These studies essentially evaluate the security of pairing wearable devices and the (lack of security) in the communication channels, e.g., due to misconfiguration or transmission of health data in plain text.
Such studies help evaluate widely used devices in the market and wireless technologies (e.g., Fitbit, Bluetooth and NFC), demonstrating security attacks and potential data breaches.
Here we also see that an integrated framework can be developed for security and privacy wearables and IoT devices.

\subsubsection{Quality of Evaluation Studies}
Overall, the quality of the reported evaluations for the existing solutions achieved a good mark of $8.57$ (out 11) as shown in Table \ref{tab:qual-eval}.
In fact, six evaluation studies achieved 10 marks, qualifying them among the top ranked studies (i.e., [ER19, ER34, ER39, ER47, ER33, ER44]).
These studies share a clear presentation addressing all the quality criteria.
These studies clearly establish the research aims and context, detail the methodology and carry out the evaluation, describing all steps from the study design to the data collection and analysis.

To put things in the perspective, four of these studies focused on evaluating privacy policies of mHealth apps ([ER19, ER34, ER39, ER47]).
They carefully describe their research plan and gather a significant number of mHealth apps and their privacy policies for evaluation (ranging from 64 [ER19] up to 600 apps [ER47]).
The evaluation follows a very structured and replicable review process, usually checking if the privacy policies are available, their readability, the quality of the content and compliance with the existing regulations.
Likewise, in [ER33] describes a server-focused security analysis, comprising 60 apps that were observed to communicate with 823 servers.
The servers were then analysed using a set of security tools (i.e., BProxy, testssl script and Qualys SSL Labs), also enabling a replicable collection and analysis of data.
The last study [ER44], evaluates the security of heartbeat-based mechanisms (biometric authentication) against a specific attack (i.e., remote photoplethysmography).
Six subjects participated in the study, enabling the collection of video sequences from which individual heartbeats were detected correctly and accurately from a distance, proving the attack feasibility.

\subsubsection{State of the Evaluation Studies}
In this systematic review we found that out of 350+ papers that proposed a new solutions in the area, a vast majority, 173, of the papers are limited to a validation study.
There were only 52 studies that performed evaluation in practice; there were only 28 studies that focused on the security and privacy solutions for m/uHealth (e.g., [ER04, ER11, ER18]).
This suggests that the empirical evidence is rather limited and most of the proposed solutions, that have only been validated, still need to be properly evaluated.

Almost all evaluation studies concentrate in the area of mHealth apps.
A small portion of the evaluation studies addressed security and privacy for more advanced uHealth systems, such as:
(a) on wearables, Internet of Things (IoT) and Internet of Medical Things (IoMT) [ER26, ER23, ER35];
(b) the use of devices and standards on medical sensor networks [ER48, ER18, ER16]; and,
(c) working with biometric sensors for authentication [ER44].

In the field of mHealth systems there has been a greater emphasises on two specific areas, i.e., (a) security and privacy for mHealth apps (e.g., [ER19, ER20, ER27]), and (b) assessments of mHealth apps' privacy policies (e.g., [ER01, ER34]). It is imperative to further develop the proposals in these areas.
Both groups of studies can be additionally analysed and compared, so as to account for multiple dimensions of security and privacy, towards an integrated framework for assessing mHealth systems.

\section{Threats to Validity of the SMS}
\label{sec:threats-to-validity}
We now present some threats to the validity of this mapping study. 
As in Figure \ref{fig:methodology} (Section \ref{sec:methodology}), we followed the guidelines and recommendations to conduct the mapping study from \cite{petersen2008systematic} that provides a systematic and objective manner to plan, conduct and document the SMS. 
However, customizations to the pre-defined methodological steps may lead to some bias and limitations that represent threats to the validity of this SMS, as detailed below. %

\subsection{Threat I -- limitations of the SMS plan}
The first threat relates to the planning of the SMS in terms of identifying the needs and justifications for such a mapping study. 
Considering the number of the existing secondary studies (as in Table \ref{tab:comp-surveys}), during the planning phase, there is a need to outline the scope of the mapping study that does not overlap the existing research contributions. 
In order to avoid the risk of overlapping scope, we executed the search string (Listing \ref{lst:search-string-slrs}) to ensure that there do not exist any secondary studies on a similar topic. 
The results of the search string (Section \ref{sec:identify-need}) did not return any relevant secondary study that focuses on the security and privacy of m/uHealth systems.  
Another important aspect of planning an SMS is to outline the research questions that provide the basis for an objective investigation of the studies being reviewed in the SMS. 
If the RQs are not explicitly stated or omit the key topics, the results of a mapping study can be flawed, overlooking the key information. 
In order to avoid this threat, we outlined a number of RQs and objective for each of the RQ (Section \ref{sec:research-questions}) that aim to find answers about the frequency, progression, existing challenges, solution and emerging research. 
We tried our best to minimize any bias or limitations during the planning phase to define the scope and objectives of the SMS. 
Once the SMS plan was created, it was cross checked (e.g., review of existing secondary studies, refinement of RQs) independently in an effort to minimize the limitations of the SMS plan before proceeding to the next phases.

\subsection{Threat II -- credibility of the literature search process}
After SMS planning, the next threat relates to the identification and selection of the literature that is selected for review in an SMS. 
The process of selecting the papers to be reviewed is a critical step as if the relevant papers are missed, the results of a SMS may be flawed. 
We followed a two-steps process (Section \ref{sec:study-inclusion}), referred as literature screening and qualitative assessment, to minimize the threats to the selection process of the reviewed papers in this SMS. 
Also, this SMS restricts the selection of publications to one scientific database, i.e., Scopus. 
This decision was made due to the sheer volume of publications retrieved from just one source. 
Scopus was nonetheless chosen since the previous systematic reviews have shown that the searches in this platform alone result in a far reaching number of papers that would otherwise just be duplicated papers if other databases such as IEEE Xplore, ACM and PubMed are being searched independently \cite{dieste2007search, garousi2016citations, mourao2017investigating}. 
Furthermore, the mapping studies are extendable and other scientific databases could be considered provided that the other researchers have enough resources to do so. 
In this SMS, we also narrowed the search to a five-year period (i.e., 2015-2019) in order to concentrate on the state-of-the-art in science as well as to keep the study viable. 
Based on a step-wise search process, we are confident that we have tried to minimise the limitations related to (i) excluding or overlooking any relevant study of (ii) including the irrelevant or low quality study that can impact the results and their documentation in a SMS.

\subsection{Threat III -- potential bias in SMS documentation}
The last threat relates to the potential bias in synthesising the data from the review and documenting the results.
This means if there are some limitations in the data synthesis they have a direct impact on the results of this SMS. 
Typical examples of such limitations could be flawed research taxonomy, incorrect identification of research themes (identified challenges and proposed solutions) and mismatch of emerging research trends.
In order to minimise the bias in synthesising and documenting the results, we have created the taxonomies and the research facets (detailed in our research protocol \cite{iwaya2020protocol}).
There were two researchers involved in synthesising the results for which the extracted data and synthesis were cross checked by an independent researcher in order to ensure the consistency.
Apart from that, this SMS also offers a complete replication package that conveniently enables other researchers to reproduce and/or extend this review (described in Section \ref{sec:phase-iii}).

\section{Conclusion}
\label{sec:conclusion}
The seminal work on ubiquitous computing by Mark Weiser \cite{weiser1999ubicomp} proposed that: \textit{``The most profound technologies are those that disappear. They weave themselves into the fabric of everyday life until they are indistinguishable from it.''}
Today, m/uHealth technologies carry Weiser's vision towards a pervasive and ubiquitous healthcare, promising access to health services anywhere and anytime.
In the past decade countless m/uHealth initiatives have been reported across nations of all income levels.
The provision of health services through mobile and ubiquitous devices has started revolutionising the health care systems across the globe. For example, the high income countries are leveraging the sophisticated remote patient monitoring systems, with real-time analytics and emergency response facilities; the low and middle income countries are empowering front-line workers with mHealth systems for treatment adherence and public health surveillance in unserved and under-served communities.

However, the increasing number of security attacks and data privacy compromises are proving to be the barriers to full adoption of m/uHealth systems. Researchers have been producing a significant amount of solutions, practices, tools and guidelines to address the security and privacy challenges of m/uHealth systems.
This study has systematically selected, analyzed and synthesized the relevant literature on the security and privacy of m/uHealth systems using an evidence-based software engineering methodology, systematic mapping study. The results of this study are expected to be beneficial and insightful for the research community, developers and practitioners to quickly figure out and understand the security and privacy challenges of m/uHealth systems and to determine the areas of future research.

\textit{Contributions and implications of the SMS:} This SMS has identified and mapped the most of the research on security and privacy of m/uHealth systems using the technical controls for safeguarding health data.
Far less attention is given to other non-technical and organisational controls, even though they are critical for every organisation.
Most part of the research papers also propose new solutions that are only validated, instead of evaluated in practice.
That indicates that more rigorous evaluation processes need to be adopted by researchers in order to strengthen the scientific evidence, and thus, foster widespread adoption of security and privacy solutions in the context of m/uHealth systems.

The key contributions of this SMS are:
\begin{itemize}
    \item Classify and compare the existing and emerging solutions, challenges and trends for security and privacy for m/uHealth.
    \item Provide an evaluation focused analysis of the solutions -- implemented in practice -- to identify commons themes and appraising the quality of these evaluation studies.
\end{itemize}

The findings of this SMS can help:
\begin{itemize}
    \item Researchers who are interested in a quick identification of the existing research.
    \item Practitioners who want to understand academic solutions that could be adopted in an industrial context.
\end{itemize}

\textit{Needs for futuristic systematic reviews:} For future work, additional systematic reviews in the field would be beneficial, whether in the form of systematic literature reviews or mapping studies.
For instance, literature reviews could focus on specific control families (e.g., focus solely on individual participation), types of application or technologies.
Literature reviews or mapping studies can also be conducted for over-represented areas, since they can be narrowed-down and further analysed (e.g., if feasible, perhaps at a control-level instead of a family-level).

\appendices
\section{Complete list of primary studies}
\label{app:primary-studies-list}
\scriptsize
[P1]	A privacy protection for an mHealth messaging system.	2015.

[P2]	OWASP inspired mobile security.	2015.	

[P3]	Watermarking of Parkinson disease speech in cloud-based healthcare framework.	2015.	

[P4]	CPLM: Cloud facilitated privacy shielding leakage resilient mobile health monitoring.	2015.	

[P5]	Anytime, anywhere access to secure, privacy-aware healthcare services: Issues, approaches and challenges.	2015.		

[P6]	Systematic information flow control in mHealth systems.	2015.

[P7]	RSA-DABE: A novel approach for secure health data sharing in ubiquitous computing environment.	2015.

[P8]	Security and privacy issues in implantable medical devices: A comprehensive survey.	2015.	

[P9]	Security aspects of cloud based mobile health care application.	2015.

[P10]	Power consumption aware software architecture for M-health applications with adaptive security of network protocols.	2015.

[P11]	Securing XML with role-based access control: Case study in health care.	2015.

[P12]	Reducing energy consumption of mobile phones during data transmission and encryption for wireless body area network applications.	2015.	

[P13]	Secure M-health patient monitoring and emergency alert system framework.	2015.

[P14]	Privacy issues and techniques in e-health systems.	2015.			

[P15]	Verifiable privacy-preserving monitoring for cloud-assisted mHealth systems.	2015.

[P16]	A Lightweight Encryption Scheme Combined with Trust Management for Privacy-Preserving in Body Sensor Networks.	2015.	

[P17]	Secu Wear: An Open Source, Multi-component Hardware/Software Platform for Exploring Wearable Security.	2015.

[P18]	Lightweight and privacy-preserving agent data transmission for mobile Healthcare.	2015.

[P19]	EPPS: Efficient and privacy-preserving personal health information sharing in mobile healthcare social networks.	2015.

[P20]	MHealth through quantified-self: A user study.	2015.

[P21]	Novel key management for secure information of ubiquitous healthcare domains to APT attack.	2015.

[P22]	Security of personal bio data in mobile health applications for the elderly.	2015.

[P23]	Security testing for Android mHealth apps.	2015.

[P24]	On the privacy, security and safety of blood pressure and diabetes apps.	2015.	

[P25]	Service intelligence and communication security for ambient assisted living.	2015.	

[P26]	Privacy-preserving mobile access to Personal Health Records through Google's Android.	2015.

[P27]	Legal, Regulatory, and Risk Management Issues in the Use of Technology to Deliver Mental Health Care.	2015.	

[P28]	Privacy preserving classification of ECG signals in mobile e-health applications.	2015.

[P29]	An Elliptic Curve Cryptography-Based RFID Authentication Securing E-Health System.	2015.

[P30]	An effective and secure user authentication and key agreement scheme in m-healthcare systems.	2015.

[P31]	Recommendation-based trust management in body area networks for mobile healthcare.	2015.

[P32]	Privacy and Security in Mobile Health Apps: A Review and Recommendations.	2015.

[P33]	Exploring mobile health in a private online social network.	2015.

[P34]	Too Much Information: Visual Research Ethics in the Age of Wearable Cameras.	2015.

[P35]	Health Care Providers' Perspectives on a Weekly Text-Messaging Intervention to Engage HIV-Positive Persons in Care (WelTel BC1).	2015.

[P36]	Development of mHealth system for supporting self-management and remote consultation of skincare eHealth/ telehealth/ mobile health systems.	2015.

[P37]	BlinkToSCoAP: An end-to-end security framework for the Internet of Things.	2015.

[P38]	A Taxonomy of mHealth apps - Security and privacy concerns.	2015.

[P39]	Designing for scalability and trustworthiness in mHealth systems.	2015.

[P40]	An Energy Efficient Method for Secure and Reliable Data Transmission in Wireless Body Area Networks Using RelAODV.	2015.

[P41]	'The phone reminder is important, but will others get to know about my illness?' Patient perceptions of an mHealth antiretroviral treatment support intervention in the HIVIND trial in South India.	2015.

[P42]	Citizen Science on Your Smartphone: An ELSI Research Agenda.	2015.

[P43]	A novel decentralized trust evaluation model for secure mobile healthcare systems.	2015.

[P44]	Polynomial based light weight security in wireless body area network.	2015.

[P45]	SmartHealth-NDNoT: Named data network of things for healthcare services.	2015.

[P46]	Trust, Perceived Risk, Perceived Ease of Use and Perceived Usefulness as Factors Related to mHealth Technology Use.	2015.

[P47]	A review and comparative analysis of security risks and safety measures of mobile health apps.	2015.	

[P48]	How trustworthy are apps for maternal and child health?.	2015.

[P49]	On using a von neumann extractor in heart-beat-based security.	2015.

[P50]	Know your audience: Predictors of success for a patient-centered texting app to augment linkage to HIV care in rural Uganda.	2015.

[P51]	SecourHealth: A delay-tolerant security framework for mobile health data collection.	2015.

[P52]	Availability and quality of mobile health app privacy policies.	2015.

[P53]	An open-access mobile compatible electronic patient register for rheumatic heart disease ('eRegister') based on the World Heart Federation's framework for patient registers.	2015.	

[P54]	Mobile early detection and connected intervention to coproduce better care in severe mental illness.	2015.		

[P55]	'Trust but verify' - five approaches to ensure safe medical apps.	2015.

[P56]	A framework for secured collaboration in mHealth.	2015.

[P57]	Security and privacy for mobile healthcare networks: From a quality of protection perspective.	2015.

[P58]	Security and privacy framework for ubiquitous healthcare IoT devices.	2016.

[P59]	Security in Cloud-Computing-Based Mobile Health.	2016.

[P60]	Mobile health (m-Health) system in the context of IoT.	2016.

[P61]	A Telemonitoring Framework for Android Devices.	2016.

[P62]	Real-Time Tele-Monitoring of Patients with Chronic Heart-Failure Using a Smartphone: Lessons Learned.	2016.

[P63]	Mobile admittance of health information with privacy and analysis in telemedicine.	2016.

[P64]	Preserving patient's anonymity for mobile healthcare system in IoT environment.	2016.

[P65]	Patient monitoring system for cardiovascular disease based on smart mobile healthcare environments.	2016.

[P66]	A trust-based framework for information sharing between mobile health care applications.	2016.		

[P67]	An information privacy risk index for mHealth apps.	2016.

[P68]	Realising the technological promise of smartphones in addiction research and treatment: An ethical review.	2016.	

[P69]	PMHM: Privacy in Mobile Health Monitoring using identity based encryption for mHealth: A Paper for mHealth systems.	2016.

[P70]	Improving security in portable medical devices and mobile health care system using trust.	2016.	

[P71]	A smart health application and its related privacy issues.	2016.

[P72]	An efficient MAC-based scheme against pollution attacks in XOR network coding-enabled WBANs for remote patient monitoring systems.	2016.

[P73]	Flexible authentication protocol with key reconstruction in WBAN environments.	2016.

[P74]	On the security of "verifiable privacy-preserving monitoring for cloud-assisted mhealth systems".	2016.		

[P75]	Private Data Analytics on Biomedical Sensing Data via Distributed Computation.	2016.

[P76]	Privacy-preserving mhealth data release with pattern consistency.	2016.

[P77]	Efficient attribute-based secure data sharing with hidden policies and traceability in mobile health networks.	2016.

[P78]	A bilinear pairing based anonymous authentication scheme in wireless body area networks for mHealth.	2016.

[P79]	Ethical guidelines for mobile app development within health and mental health fields.	2016.	

[P80]	Authentication and key relay in medical cyber-physical systems.	2016.

[P81]	The extent to which the POPI act makes provision for patient privacy in mobile personal health record systems.	2016.

[P82]	An approach to automate health monitoring in compliance with personal privacy.	2016.		

[P83]	An intelligent RFID-enabled authentication scheme for healthcare applications in vehicular mobile cloud.	2016.

[P84]	Providing security and fault tolerance in P2P connections between clouds for mHealth services.	2016.	

[P85]	Assessing pairing and data exchange mechanism security in the wearable internet of things.	2016.

[P86]	Security enhancement on an authentication scheme for privacy preservation in Ubiquitous Healthcare System.	2016.

[P87]	Attribute-based traceable anonymous proxy signature strategy for mobile healthcare.	2016.		

[P88]	Open source based privacy-proxy to restrain connectivity of mobile apps.	2016.		

[P89]	Analyzing privacy risks of mhealth applications.	2016.

[P90]	Simulation environment for testing security and privacy of mobile health apps.	2016.

[P91]	Security Recommendations for mHealth Apps: Elaboration of a Developer's Guide.	2016.

[P92]	Integration of smart wearable mobile devices and cloud computing in South African healthcare.	2016.

[P93]	Power attack: An emerging threat in health-care applications using medical body area networks.	2016.

[P94]	Secure and privacy preserving mobile healthcare data exchange using cloud service.	2016.		

[P95]	Privacy challenges and goals in mHealth systems.	2016.

[P96]	Efficient data integrity auditing for storage security in mobile health cloud.	2016.

[P97]	Analysis of ISO/IEEE 11073 built-in security and its potential IHE-based extensibility.	2016.	

[P98]	Privacy, Trust and Security in Two-Sided Markets.	2016.

[P99]	Georeferenced and secure mobile health system for large scale data collection in primary care.	2016.	

[P100]	Resolving conflicting privacy policies in M-health based on prioritization.	2016.

[P101]	mSieve: Differential behavioral privacy in time series of mobile sensor data.	2016.		

[P102]	Mutual authentication protocol for secure NFC based mobile healthcard.	2016.		

[P103]	A spatio-situation-based access control model for dynamic permission on mobile applications.	2016.			

[P104]	Harnessing teams and technology to improve outcomes in infants with single ventricle.	2016.

[P105]	Improved population health surveillance and chronic disease management using secure email: Application of the DIRECT, IEEE 11073, HITSP, and IHE standards and protocols.	2016.

[P106]	Lightweight and confidential data aggregation in healthcare wireless sensor networks.	2016.	

[P107]	Secure candy castle - A prototype for privacy-aware mHealth apps.	2016.

[P108]	Security analysis of the IEEE 802.15.6 standard.	2016.

[P109]	Context-aware, knowledge-intensive, and patient-centric Mobile Health Care Model.	2016.

[P110]	Analysis of Security Protocols for Mobile Healthcare.	2016.

[P111]	The introduction and evaluation of mobile devices to improve access to patient records: A catalyst for innovation and collaboration.	2016.

[P112]	U-prove based security framework for mobile device authentication in eHealth networks.	2016.

[P113]	Sharing mHealth data via named data networking.	2016.			

[P114]	A Secure System for Pervasive Social Network-Based Healthcare.	2016.

[P115]	On the Deployment of Healthcare Applications over Fog Computing Infrastructure.	2017.

[P116]	Information classification scheme for next generation access control models in mobile patient-centered care systems.	2017.			

[P117]	Implementation of energy efficient/lightweight encryption algorithm for wireless body area networks.	2017.

[P118]	Mapping Security Requirements of Mobile Health Systems into Software Development Lifecycle.	2017.

[P119]	Exploring the usability, security and privacy taxonomy for mobile health applications.	2017.

[P120]	An access control framework for secure and interoperable cloud computing applied to the healthcare domain.	2017.			

[P121]	A provably secure RFID authentication protocol based on elliptic curve signature with message recovery suitable for m-Health environments.	2017.

[P122]	New watermarking/encryption method for medical images full protection in m-Health.	2017.

[P123]	Smartphone application for medical images secured exchange based on encryption using the matrix product and the exclusive addition.	2017.

[P124]	Towards secure and flexible EHR sharing in mobile health cloud under static assumptions.	2017.

[P125]	A compressive multi-kernel method for privacy-preserving machine learning.	2017.

[P126]	Managing secure personal mobile health information.	2017.

[P127]	Security analysis of a mHealth app in Android: Problems and solutions.	2017.	

[P128]	Anonymity-preserving methods for client-side filtering in position-based collaboration approaches.	2017.

[P129]	Smartphones to access to patient data in hospital settings: Authentication solutions for shared devices.	2017.

[P130]	An authentication scheme for wireless healthcare monitoring sensor network.	2017.		

[P131]	Improving the information security of personal electronic health records to protect a patient's health information.	2017.

[P132]	A secure framework for mHealth data analytics with visualization.	2017.

[P133]	A proposal to improve the authentication process in m-health environments.	2017.

[P134]	Feasibility of a secure wireless sensing smartwatch application for the self-management of pediatric asthma.	2017.

[P135]	Smartphone-based continuous blood pressure monitoring application-robust security and privacy framework.	2017.

[P136]	Security of ePrescriptions: Data in transit comparison using existing and mobile device services.	2017.

[P137]	Efficient privacy-preserving dot-product computation for mobile big data.	2017.

[P138]	A privacy-preserving data sharing solution for mobile healthcare.	2017.

[P139]	Private and Secured Medical Data Transmission and Analysis for Wireless Sensing Healthcare System.	2017.

[P140]	Secure and Privacy-Preserving Data Sharing and Collaboration in Mobile Healthcare Social Networks of Smart Cities.	2017.

[P141]	Preserving patients’ privacy in health scenarios through a multicontext-aware system.	2017.	

[P142]	Efficient end-to-end authentication protocol for wearable health monitoring systems.	2017.		

[P143]	Exploring the need for a suitable privacy framework for mHealth when managing chronic diseases.	2017.

[P144]	Lightweight trusted authentication protocol for Wireless Sensor Network in e-Health.	2017.

[P145]	Security and privacy requirements engineering for human centric IoT systems using eFRIEND and Isabelle.	2017.

[P146]	The t2rol access control model for mobile health systems in developing countries.	2017.		

[P147]	Privacy Requirements for mobile e-Service in the Health Authority-Abu Dhabi (HAAD).	2017.

[P148]	Protection against wormhole attack using smart protocol in MANET.	2017.

[P149]	Cybersecurity of Wearable Devices: An Experimental Analysis and a Vulnerability Assessment Method.	2017.

[P150]	Trusted sensor signal protection for confidential point-of-care medical diagnostic	2017.

[P151]	Effective? Engaging? secure? applying the orcha-24 framework to evaluate apps for chronic insomnia disorder.	2017.

[P152]  Development of a web-based epidemiological surveillance system with health system response for improving maternal and newborn health: Field-testing in Thailand.	2017.

[P153]	Protecting mobile health records in cloud computing: A secure, efficient, and anonymous design.	2017.

[P154]	A collaborative privacy-preserving deep learning system in distributed mobile environment.	2017.

[P155]	Secure authentication and Prescription Safety Protocol for telecare health services using ubiquitous IoT.	2017.

[P156]	Developing countries and Internet-of-Everything (IoE).	2017.

[P157]	A Privacy-Preserving Multi-Authority Attribute-Based Encryption Approach for Mobile Healthcare.	2017.

[P158]	PlaIMoS: A remote mobile healthcare platform to monitor cardiovascular and respiratory variables.	2017.

[P159]	SoTRAACE - Socio-technical risk-adaptable access control model.	2017.		

[P160]	Pseudonym-based privacy protection for Steppy application.	2017.

[P161]	My smart age with HIV: An innovative mobile and IoMT framework for patient's empowerment.	2017.

[P162]	A survey on IoT applications, security challenges and counter measures.	2017.

[P163]	Using Attribute-Based Access Control for Remote Healthcare Monitoring.	2017.

[P164]	Accelerometer and Fuzzy Vault-Based Secure Group Key Generation and Sharing Protocol for Smart Wearables.	2017.

[P165]	Patients’ data management system protected by identity-based authentication and key exchange.	2017.

[P166]	An Efficient Activity Recognition Framework: Toward Privacy-Sensitive Health Data Sensing.	2017.

[P167]	Secure internet of things-based cloud framework to control zika virus outbreak.	2017.

[P168]	Enhancing Heart-Beat-Based Security for mHealth Applications.	2017.

[P169]	ABE based raspberry pi secure health sensor (SHS).	2017.

[P170]	Review on communication security issues in iot medical devices.	2017.

[P171]	A multimodal biometric identification method for mobile applications security.	2017.

[P172]	Extended provisioning, security and analysis techniques for the ECHO health data management system.	2017.

[P173]	Trustworthy access control for wireless body area networks.	2017.

[P174]	Utilizing fully homomorphic encryption to implement secure medical computation in smart cities.	2017.

[P175]	The new wave of privacy concerns in the wearable devices era.	2017.

[P176]	Toward improving electrocardiogram (ECG) biometric verification using mobile sensors: A two-stage classifier approach.	2017.

[P177]	A security framework for mobile health applications.	2017.

[P178]	Optimized and Secured Transmission and Retrieval of Vital Signs from Remote Devices.	2017.

[P179]	Soft real-time smartphone ECG processing.	2017.

[P180]	An efficient secure communication for healthcare system using wearable devices.	2017.

[P181]	Developing a comprehensive information security framework for mHealth: a detailed analysis.	2017.

[P182]	Blockchain as an enabler for public mHealth solutions in South Africa.	2017.

[P183]	Secure-EPCIS: Addressing Security Issues in EPCIS for IoT Applications.	2017.

[P184]	The design of an m-Health monitoring system based on a cloud computing platform.	2017.	

[P185]	Lightweight Sharable and Traceable Secure Mobile Health System.	2017.

[P186]	mHealth quality: A process to seal the qualified mobile health apps.	2017.

[P187]	Scalable Role-Based Data Disclosure Control for the Internet of Things.	2017.

[P188]	Context aware based user customized light therapy service using security framework.	2017.

[P189]	Privacy-preserving design for emergency response scheduling system in medical social networks.	2017.

[P190]	Distributed Big Data Analytics in Service Computing.	2017.

[P191]	Light-Weight and Robust Security-Aware D2D-Assist Data Transmission Protocol for Mobile-Health Systems.	2017.

[P192]	DFP: A data fragment protection scheme for mhealth in wireless network.	2017.

[P193]	Anonymous anti-sybil attack protocol for mobile healthcare networks analytics.	2017.

[P194]	Towards privacy protection and malicious behavior traceability in smart health.	2017.

[P195]	Attribute-based encryption with non-monotonic access structures supporting fine-grained attribute revocation in m-healthcare.	2017.

[P196]	Secure framework for patient data transmission on mobile-cloud platform.	2018.

[P197]	A Secure Crypto Base Authentication and Communication Suite in Wireless Body Area Network (WBAN) for IoT Applications.	2018.	

[P198]	Designing a secure architecture for m-health applications.	2018.

[P199]	Towards an architecture to guarantee both data privacy and utility in the first phases of digital clinical trials.	2018.

[P200]	A stochastic game for adaptive security in constrained wireless body area networks.	2018.

[P201]	Context-aware authorization and anonymous authentication in wireless body area networks.	2018.

[P202]	Game-Based Adaptive Remote Access VPN for IoT: Application to e-Health.	2018.

[P203]	Spatially-resolved estimation of personal dosage of airborne particulates for ambulatory subjects using wearable sensors.	2018.		

[P204]	A Physical Layer Security Scheme for Mobile Health Cyber-Physical Systems.	2018.

[P205]	Evaluating the Privacy Policies of Mobile Personal Health Records for Pregnancy Monitoring.	2018.

[P206]	An efficient implementation of next generation access control for the mobile health cloud.	2018.

[P207]	Secure sharing of mobile personal healthcare records using certificateless proxy re-encryption in cloud.	2018.

[P208]	Quality of publicly available physical activity apps: Review and content analysis.	2018.

[P209]	Using Mobile Phone Sensor Technology for Mental Health Research: Integrated Analysis to Identify Hidden Challenges and Potential Solutions.	2018.

[P210]	Secured cloud computing for medical data based on watermarking and encryption.	2018.

[P211]	Are mHealth Apps Secure? A Case Study.	2018.

[P212]	Challenges and solutions implementing an SMS text message-based survey CASI and adherence reminders in an international biomedical HIV PrEP study (MTN 017).	2018.	

[P213]	Threat modeling for mobile health systems	2018.		

[P214]	RoFa: A Robust and Flexible Fine-Grained Access Control Scheme for Mobile Cloud and IoT based Medical Monitoring.	2018.

[P215]	Verifiable keyword search for secure big data-based mobile healthcare networks with fine-grained authorization control.	2018.	

[P216]	A real-time, automated and privacy-preserving mobile emergency-medical-service network for informing the closest rescuer to rapidly support mobile-emergency-call victims.	2018.

[P217]	A Simple Approach for Securing IoT Data Transmitted over Multi-RATs.	2018.

[P218]	On the cybersecurity of m-Health IoT systems with LED bitslice implementation.	2018.		

[P219]	mHealth applications for goal management training - Privacy engineering in neuropsychological studies.	2018.

[P220]	A Secured Smartphone-Based Architecture for Prolonged Monitoring of Neurological Gait.	2018.		

[P221]	A patient-held smartcard with a unique identifier and an mhealth platform to improve the availability of prenatal test results in rural Nigeria: Demonstration study.	2018.

[P222]	Using mobile location data in biomedical research while preserving privacy.	2018.

[P223]	Evaluating authentication options for mobile health applications in younger and older adults.	2018.

[P224]	Comparative analysis between different facial authentication tools for assessing their integration in m-health mobile applications.	2018.

[P225]	A Provably-Secure Cross-Domain Handshake Scheme with Symptoms-Matching for Mobile Healthcare Social Network.	2018.

[P226]	Secure identity-based data sharing and profile matching for mobile healthcare social networks in cloud computing.	2018.	

[P227]	Privacy preserved spectral analysis using iot mhealth biomedical data for stress estimation.	2018.

[P228]	A security framework for mHealth apps on Android platform.	2018.

[P229]	Assessing the privacy of mhealth apps for self-tracking: Heuristic evaluation approach.	2018.

[P230]	Skin care management support system based on cloud computing.	2018.

[P231]	Toward privacy-preserving symptoms matching in SDN-based mobile healthcare social networks.	2018.

[P232]	Securing Medical Images for Mobile Health Systems Using a Combined Approach of Encryption and Steganography.	2018.

[P233]	Toward privacy in IoT mobile devices for activity recognition.	2018.

[P234]	New Engineering Method for the Risk Assessment: Case Study Signal Jamming of the M-Health Networks.	2018.

[P235]	Smart Chair-A Telemedicine Based Health Monitoring System.	2018.

[P236]	Cloud-assisted mutual authentication and privacy preservation protocol for telecare medical information systems.	2018.		

[P237]	Cost-Effective and Anonymous Access Control for Wireless Body Area Networks.	2018.

[P238]	Secure and efficient anonymous authentication scheme for three-tier mobile healthcare systems with wearable sensors.	2018.	

[P239]	Towards blockchain empowered trusted and accountable data sharing and collaboration in mobile healthcare applications.	2018.

[P240]	ETAP: Energy-Efficient and Traceable Authentication Protocol in Mobile Medical Cloud Architecture.	2018.	

[P241]	Secure and efficient querying over personal health records in cloud computing.	2018.	

[P242]	Attribute-based handshake protocol for mobile healthcare social networks.	2018.

[P243]	A traceable threshold attribute-based signcryption for mHealthcare social network.	2018.

[P244]	Efficient Fine-Grained Data Sharing Mechanism for Electronic Medical Record Systems with Mobile Devices.	2018.		

[P245]	Certificateless searchable public key encryption scheme for mobile healthcare system.	2018.	

[P246]	A vulnerability study of Mhealth chronic disease management (CDM) applications (apps).	2018.		

[P247]	A design thinking approach to implementing an android biometric unique identification system for infant treatment follow-up in a resource limited setting.	2018.

[P248]	My Smart Remote: A Smart Home Management Solution for Children.	2018.

[P249]	Detecting insider attacks in medical cyber-physical networks based on behavioral profiling.	2018.	

[P250]	Privacy Issues in Smartphone Applications: An Analysis of Headache/Migraine Applications.	2018.

[P251]	An authentication protocol for smartphone integrated Ambient Assisted Living system.	2018.

[P252]	Secure and lightweight remote patient authentication scheme with biometric inputs for mobile healthcare environments.	2018.

[P253]	Real-Time Remote Health Monitoring Systems Using Body Sensor Information and Finger Vein Biometric Verification: A Multi-Layer Systematic Review.	2018.

[P254]	OpSecure: A secure unidirectional optical channel for implantable medical devices.	2018.

[P255]	Translating GDPR into the mHealth Practice.	2018.

[P256]	Security and Privacy Analysis of Mobile Health Applications: The Alarming State of Practice.	2018.

[P257]	HR-auth: Heart rate data authentication using consumer wearables.	2018.	

[P258]	Privacy Issues and Solutions for Consumer Wearables.	2018.

[P259]	An Improved Asymmetric Key Based Security Architecture for WSN.	2018.

[P260]	Spatial Blockchain-Based Secure Mass Screening Framework for Children with Dyslexia.	2018.

[P261]	Exploiting smart e-Health gateways at the edge of healthcare Internet-of-Things: A fog computing approach.	2018.

[P262]	A service oriented healthcare architecture (SOHA-CC) based on cloud computing.	2018.

[P263]	Chaotic map-based anonymous user authentication scheme with user biometrics and fuzzy extractor for crowdsourcing internet of things.	2018.

[P264]	Implementation of cloud-assisted secure data transmission in WBAN for healthcare monitoring.	2018.		

[P265]	Dynamic Connectivity Establishment and Cooperative Scheduling for QoS-Aware Wireless Body Area Networks.	2018.

[P266]	SMEAD: A secured mobile enabled assisting device for diabetics monitoring.	2018.

[P267]	How could commercial terms of use and privacy policies undermine informed consent in the age of mobile health?.	2018.

[P268]	Assessment criteria for parents to determine the trustworthiness of maternal and child health apps: a pilot study.	2018.

[P269]	Attacks on heartbeat-based security using remote photoplethysmography.	2018.

[P270]	ECG-Based Secure Healthcare Monitoring System in Body Area Networks.	2018.

[P271]	A public-key protection scheme using in the wireless module of the digital stethoscope.	2018.

[P272]	Centralized Fog Computing security platform for IoT and cloud in healthcare system.	2018.			

[P273]	FoG assisted secure De-duplicated data dissemination in smart healthcare IoT.	2018.

[P274]	A cyber-physical approach to trustworthy operation of health monitoring systems.	2018.	

[P275]	Tiger hash based AdaBoost machine learning classifier for secured multicasting in mobile healthcare system.	2018.		

[P276]	ACMHS: Efficient access control for mobile health care system.	2018.

[P277]		P3ASC: Privacy-preserving pseudonym and attribute-based signcryption scheme for cloud-based mobile healthcare system.	2018.

[P278]		Identity-Based Fast Authentication Scheme for Smart Mobile Devices in Body Area Networks.	2018.

[P279]	Enabling efficient and privacy-preserving health query over outsourced cloud.	2018.

[P280]	EPSMD: An Efficient Privacy-Preserving Sensor Data Monitoring and Online Diagnosis System.	2018.

[P281]	Development of a just-in-time adaptive mhealth intervention for insomnia: Usability study.	2018.

[P282]	A lightweight and robust two-factor authentication scheme for personalized healthcare systems using wireless medical sensor networks.	2018.

[P283]	Securing mobile healthcare data: A smart card based cancelable Finger-Vein Bio-Cryptosystem.	2018.	

[P284]	Big Data and IoT for U-healthcare security.	2018.

[P285]	An evolutionary game-theoretic approach for assessing privacy protection in mHealth systems.	2018.

[P286]	Fusing identity management, HL7 and blockchain into a global healthcare record sharing architecture.	2019.

[P287]	Cybersecurity Metrics for Enhanced Protection of Healthcare IT Systems.	2019.

[P288]	Security vulnerabilities in mobile health applications.	2019.

[P289]	Cybersecurity and privacy issues for socially integrated mobile healthcare applications operating in a multi-cloud environment.	2019.

[P290]	The development of an Arabic weight-loss app akser waznk: Qualitative results.	2019.

[P291]	Context-aware anonymous authentication protocols in the internet of things dedicated to e-health applications.	2019.		

[P292]	Verifying a secure authentication protocol for IoT medical devices.	2019.

[P293]	What data are smartphone users willing to share with researchers?: Designing and evaluating a privacy model for mobile data collection apps.	2019.

[P294]	Privacy in mobile health applications for breast cancer patients.	2019.

[P295]	Secure application for health monitoring.	2019.

[P296]	A multivariant secure framework for smart mobile health application.	2019.

[P297]	A secure framework for IoT-based healthcare system.	2019.

[P298]	Study of out-of-hospital access to his system: A security perspective.	2019.

[P299]	Collaborative and secure transmission of medical data applied to mobile healthcare.	2019.

[P300]	A mobile phone app for bedside nursing care: Design and development using an adapted software development life cycle model.	2019.

[P301]	WYZ: A pilot study protocol for designing and developing a mobile health application for engagement in HIV care and medication adherence in youth and young adults living with HIV	2019.

[P302]	Enabling the internet of mobile crowdsourcing health things: A mobile fog computing, blockchain and iot based continuous glucose monitoring system for diabetes mellitus research and care.	2019.

[P303]	MedCop: Verifiable computation for mobile healthcare system.	2019.

[P304]	Reliable and secure data transfer in IoT networks.	2019.

[P305]	Secure sharing of mHealth data streams through cryptographically-enforced access control.	2019.

[P306]	Privacy-preserving voice-based search over mHealth data.	2019.

[P307]	Trustworthy Delegation Toward Securing Mobile Healthcare Cyber-Physical Systems.	2019.

[P308]	Data security and raw data access of contemporary mobile sensor devices.	2019.

[P309]	Review of security and privacy for the internet of medical things (IoMT): Resolving the protection concerns for the novel circular economy bioinformatics.	2019.

[P310]	CAMPS: Efficient and privacy-preserving medical primary diagnosis over outsourced cloud.	2019.			

[P311]	Mobile health systems for community-based primary care: identifying controls and mitigating privacy threats.	2019.

[P312]	E-consent for data privacy: Consent management for mobile health technologies in public health surveys and disease surveillance.	2019.

[P313]	An access control framework for protecting personal electronic health records.	2019.

[P314]	Detecting Suicidal Ideation with Data Protection in Online Communities.	2019.	

[P315]	A lightweight CLAS scheme with complete aggregation for healthcare mobile crowdsensing.	2019.	

[P316]	A multi-level data sensitivity model for mobile health data collection systems.	2019.		

[P317]	A review of privacy and usability issues in mobile health systems: Role of external factors.	2019.

[P318]	A reversible and secure patient information hiding system for IoT driven e-health.	2019.	

[P319]	A secure mutual batch authentication scheme for patient data privacy preserving in WBAN.	2019.

[P320]	Effective engagement of adolescent asthma patients with mobile health-supporting medication adherence.	2019.

[P321]	Amulet: An open-source wrist-worn platform for mHealth research and education.	2019.

[P322]	Secure IoT e-Health applications using VICINITY framework and GDPR guidelines.	2019.

[P323]	A secure elliptic curve cryptography based mutual authentication protocol for cloud-assisted TMIS.	2019.

[P324]	MSCryptoNet: Multi-Scheme Privacy-Preserving Deep Learning in Cloud Computing.	2019.

[P325]	PatientConcept App: Key Characteristics, Implementation, and its Potential Benefit.	2019.

[P326]	Fine-grained multi-authority access control in IoT-enabled mHealth.	2019.

[P327]	EdgeCare: Leveraging Edge Computing for Collaborative Data Management in Mobile Healthcare Systems.	2019.

[P328]	Lightweight and Privacy-Preserving Medical Services Access for Healthcare Cloud.	2019.

[P329]	An efficient anonymous authentication scheme based on double authentication preventing signature for mobile healthcare crowd sensing.	2019.

[P330]	Development of a security layer in a mobile health system.	2019.

[P331]	Server-Focused Security Assessment of Mobile Health Apps for Popular Mobile Platforms.	2019.

[P332]Security of an Electronic Heal Thcare System which Facilitates the Detection of Preeclampsia Through a Smart Bracelet.	2019.

[P333]	Secure and scalable mhealth data management using blockchain combined with client hashchain: System design and validation.	2019.

[P334]	Mobile apps for people with dementia: Are they compliant with the general data protection regulation (GDPR)?.	2019.		

[P335]	MHealth applications: Can user-adaptive visualization and context affect the perception of security and privacy?.	2019.	

[P336]	Eliciting design guidelines for privacy notifications in mHealth environments.	2019.

[P337]	A Novel Privacy Framework for Secure M-Health Applications: The Case of the GDPR.	2019.

[P338]	Group-Based Key Exchange for Medical IoT Device-to-Device Communication (D2D) Combining Secret Sharing and Physical Layer Key Exchange.	2019.

[P339]	Blockchain for Secure EHRs Sharing of Mobile Cloud Based E-Health Systems.	2019.

[P340]	Needs analysis for a parenting app to prevent unintentional injury in newborn babies and toddlers: Focus group and survey study among Chinese caregivers.	2019.

[P341]	Reviewing the data security and privacy policies of mobile apps for depression.	2019.

[P342]	Achieving Mobile-Health Privacy Using Attribute-Based Access Control.	2019.

[P343]	How private is your mental health app data? An empirical study of mental health app privacy policies and practices.	2019.		

[P344]	LISA: Visible light based initialization and SMS based authentication of constrained IoT devices.	2019.	

[P345]	Reliable Healthcare Monitoring System Using SPOC Framework.	2019.

[P346]	RADAR-base: Open source mobile health platform for collecting, monitoring, and analyzing data using sensors, wearables, and mobile devices.	2019.

[P347]	Availability, readability, and content of privacy policies and terms of agreements of mental health apps.	2019.

[P348]	A secure mHealth application for attention deficit and hyperactivity disorder.	2019.

[P349]	Technical guidance for clinicians interested in partnering with engineers in mobile health development and evaluation.	2019.

[P350]	Lightweight and Secure Three-Factor Authentication Scheme for Remote Patient Monitoring Using On-Body Wireless Networks.	2019.

[P351]	Sensor-Based mHealth Authentication for Real-Time Remote Healthcare Monitoring System: A Multilayer Systematic Review.	2019.

[P352]	Towards a cooperative security system for mobile-health applications.	2019.

[P353]	How to Realize Device Interoperability and Information Security in mHealth Applications.	2019.		

[P354]	End to end light weight mutual authentication scheme in IoT-based healthcare environment.	2019.			

[P355]	Robust secure communication protocol for smart healthcare system with FPGA implementation.	2019.

[P356]	Flexible and efficient authenticated key agreement scheme for BANs based on physiological features.	2019.

[P357]	Fog-Enabled Smart Health: Toward Cooperative and Secure Healthcare Service Provision.	2019.

[P358]	How to use relevant data for maximal benefit with minimal risk: Digital health data governance to protect vulnerable populations in low-income and middle-income countries.	2019.

[P359]	Achieve Privacy-Preserving Priority Classification on Patient Health Data in Remote eHealthcare System.	2019.

[P360]	An efficient and privacy-Preserving pre-clinical guide scheme for mobile eHealthcare.	2019.

[P361]	BYOD in Hospitals-Security Issues and Mitigation Strategies.	2019.

[P362]	Why reviewing apps is not enough: Transparency for trust (T4T) principles of responsible health app marketplaces.	2019.

[P363]	PDVocal: Towards privacy-preserving Parkinson's disease detection using non-speech body sounds.	2019.		

[P364]	An improved lightweight certificateless generalized signcryption scheme for mobile-health system.	2019.

[P365]	Barriers to and facilitators of the use of mobile health apps from a security perspective: Mixed-methods study.	2019.

\section{Evaluation Research list of relevant papers}
\label{app:evaluation-research-studies}
\scriptsize
[ER01] Security vulnerabilities in mobile health applications.	2019.

[ER02] The development of an Arabic weight-loss app akser waznk: Qualitative results.	2019.

[ER03] Real-Time Tele-Monitoring of Patients with Chronic Heart-Failure Using a Smartphone: Lessons Learned.	2016.

[ER04] Evaluating the Privacy Policies of Mobile Personal Health Records for Pregnancy Monitoring.	2018.

[ER05] What data are smartphone users willing to share with researchers?: Designing and evaluating a privacy model for mobile data collection apps.	2019.	

[ER06] Privacy in mobile health applications for breast cancer patients.	2019.	

[ER07] Quality of publicly available physical activity apps: Review and content analysis.	2018.

[ER08] Using Mobile Phone Sensor Technology for Mental Health Research: Integrated Analysis to Identify Hidden Challenges and Potential Solutions.	2018.	

[ER09] Challenges and solutions implementing an SMS text message-based survey CASI and adherence reminders in an international biomedical HIV PrEP study (MTN 017).	2018. 

[ER10] An information privacy risk index for mHealth apps.	2016.

[ER11] Security analysis of a mHealth app in Android: Problems and solutions.	2017.

[ER12] A mobile phone app for bedside nursing care: Design and development using an adapted software development life cycle model.	2019.

[ER13] WYZ: A pilot study protocol for designing and developing a mobile health application for engagement in HIV care and medication adherence in youth and young adults living with HIV.	2019.

[ER14] Comparative analysis between different facial authentication tools for assessing their integration in m-health mobile applications.	2018.	

[ER15] Privacy-preserving mhealth data release with pattern consistency.	2016.	

[ER16] Data security and raw data access of contemporary mobile sensor devices.	2019.

[ER17] Security of ePrescriptions: Data in transit comparison using existing and mobile device services.	2017.

[ER18] Private and Secured Medical Data Transmission and Analysis for Wireless Sensing Healthcare System.	2017.	

[ER19] Assessing the privacy of mhealth apps for self-tracking: Heuristic evaluation approach.	2018.

[ER20] Security testing for Android mHealth apps.	2015.	

[ER21] Effective engagement of adolescent asthma patients with mobile health-supporting medication adherence.	2019.

[ER22] PatientConcept App: Key Characteristics, Implementation, and its Potential Benefit.	2019.

[ER23] Cybersecurity of Wearable Devices: An Experimental Analysis and a Vulnerability Assessment Method.	2017.

[ER24] Effective? Engaging? secure? applying the orcha-24 framework to evaluate apps for chronic insomnia disorder.	2017.	

[ER25] Development of a web-based epidemiological surveillance system with health system response for improving maternal and newborn health: Field-testing in Thailand.	2017.	

[ER26] Assessing pairing and data exchange mechanism security in the wearable internet of things.	2016.	

[ER27] A vulnerability study of Mhealth chronic disease management (CDM) applications (apps).	2018.		

[ER28] Detecting insider attacks in medical cyber-physical networks based on behavioral profiling.	2018.	

[ER29] Analyzing privacy risks of mhealth applications.	2016.

[ER30] Privacy Issues in Smartphone Applications: An Analysis of Headache/Migraine Applications.	2018.

[ER31] Mobile apps for people with dementia: Are they compliant with the general data protection regulation (GDPR)?.	2019.	

[ER32] Health Care Providers' Perspectives on a Weekly Text-Messaging Intervention to Engage HIV-Positive Persons in Care (WelTel BC1).	2015.

[ER33] Server-Focused Security Assessment of Mobile Health Apps for Popular Mobile Platforms.	2019.

[ER34] Reviewing the data security and privacy policies of mobile apps for depression.	2019.

[ER35] My smart age with HIV: An innovative mobile and IoMT framework for patient's empowerment.	2017.		

[ER36] Security and Privacy Analysis of Mobile Health Applications: The Alarming State of Practice.	2018.

[ER37] How private is your mental health app data? An empirical study of mental health app privacy policies and practices.	2019.

[ER38] RADAR-base: Open source mobile health platform for collecting, monitoring, and analyzing data using sensors, wearables, and mobile devices.	2019.

[ER39] Availability, readability, and content of privacy policies and terms of agreements of mental health apps.	2019.

[ER40] 'The phone reminder is important, but will others get to know about my illness?' Patient perceptions of an mHealth antiretroviral treatment support intervention in the HIVIND trial in South India.	2015.

[ER41] Georeferenced and secure mobile health system for large scale data collection in primary care.	2016.	

[ER42] A review and comparative analysis of security risks and safety measures of mobile health apps.	2015.

[ER43] How trustworthy are apps for maternal and child health?.	2015.

[ER44] Attacks on heartbeat-based security using remote photoplethysmography.	2018.

[ER45] Harnessing teams and technology to improve outcomes in infants with single ventricle.	2016.

[ER46] Know your audience: Predictors of success for a patient-centered texting app to augment linkage to HIV care in rural Uganda.	2015.

[ER47] Availability and quality of mobile health app privacy policies.	2015.

[ER48] Security analysis of the IEEE 802.15.6 standard.	2016.

[ER49] An open-access mobile compatible electronic patient register for rheumatic heart disease ('eRegister') based on the World Heart Federation's framework for patient registers.	2015.

[ER50] Development of a just-in-time adaptive mhealth intervention for insomnia: Usability study.	2018.

[ER51] Mobile early detection and connected intervention to coproduce better care in severe mental illness.	2015.	

[ER52] The introduction and evaluation of mobile devices to improve access to patient records: A catalyst for innovation and collaboration.	2016.

\section*{Acknowledgment}
The work has been supported by the Cyber Security Research Centre Limited whose activities are partially funded by the Australian Government's Cooperative Research Centres Programme.

\bibliography{myrefs} 
\bibliographystyle{ieeetr}

\begin{IEEEbiography}[{\includegraphics[width=1in,height=1.25in,clip,keepaspectratio]{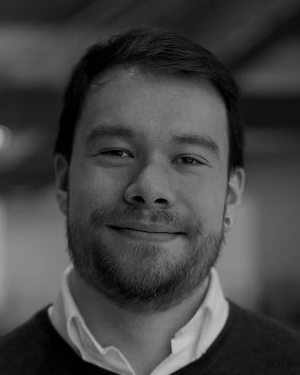}}]{Leonardo Horn Iwaya} was born in Joinville, Santa Catarina, SC, Brazil. He received the B.S. degree in computer science from the Santa Catarina State University, the M.S. degree in electrical engineering from the University of S\~{a}o Paulo, and the Ph.D. degree in computer science from Karlstad University, Sweden, in 2019. From 2011 to 2014, he was a Research Assistant with the Laboratory of Computer Networks and Architecture (LARC). Since 2019, he has been a Postdoctoral Research Fellow with the Centre for Research on Engineering Software Technologies (CREST), School of Computer Science at the University of Adelaide, Australia, in projects funded by the Cyber Security Cooperative Research Centre (CSCRC). His research interests include privacy engineering, information security, mobile and ubiquitous health systems, and the privacy impacts of new technologies.
\end{IEEEbiography}

\begin{IEEEbiography}[{\includegraphics[width=1in,height=1.25in,clip,keepaspectratio]{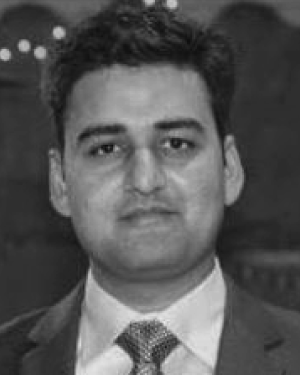}}]{Aakash Ahmad} received the Ph.D. degree in software engineering from Dublin City University, Dublin, Ireland, in 2015. He was a Postdoctoral Researcher with the IT University of Copenhagen and a Software Engineer with Elixir Technologies. He is serving as an Assistant Professor with the College of Computer Science and Engineering, University of Hail, Ha'il, Saudi Arabia. His research interests include software architecture and software patterns for mobile and cloud computing systems.
\end{IEEEbiography}

\begin{IEEEbiography}[{\includegraphics[width=1in,height=1.25in,clip,keepaspectratio]{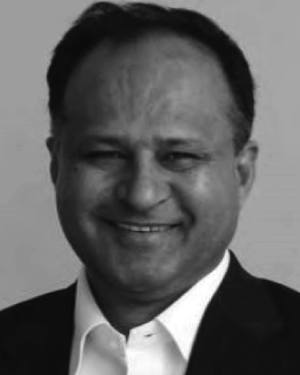}}]{M. Ali Babar} is currently a Professor with the School of Computer Science, The University of Adelaide. He is an Honorary Visiting Professor with the Software Institute, Nanjing University, China. He is the Director of Cyber Security Adelaide (CSA), which incorporates a node of recently approved the Cyber Security Cooperative Research Centre (CSCRC), whose estimated budget is around A\$140 Millions over seven years with A\$50 Millions provided by the Australia Government. In the area of Software Engineering education, he led the University's effort to redevelop a Bachelor of Engineering (software) degree that has been accredited by the Australian Computer Society and the Engineers Australia (ACS/EA). Prior to joining The University of Adelaide, he spent almost seven years in Europe (Ireland, Denmark, and U.K.) working as a Senior Researcher and an Academic. Before returning to Australia, he was a Reader of software engineering with Lancaster University. He has established an Interdisciplinary Research Centre, Centre for Research on Engineering Software Technologies (CREST), where he leads the research and research training of more than 20 (12 Ph.D. students) members. Apart from his work having industrial relevance as evidenced by several R\&D projects and setting up a number of collaborations in Australia and Europe with industry and government agencies, his publications have been highly cited within the discipline of Software Engineering as evidenced by his H-index is 48 with 8855 citations as per Google Scholar on June 20, 2020. He leads the theme on Platform and Architecture for Cyber Security as a Service with the Cyber Security Cooperative Research Centre. He has authored/coauthored more than 220 peer-reviewed publications through premier Software Technology journals and conferences.
\end{IEEEbiography}

\EOD

\end{document}